\documentclass[jcap,superscriptaddress,nofootinbib,longbibliography,amsmath,amsfonts,preprintnumbers,notitlepage,10pt,english]{revtex4-1}
\usepackage[colorlinks,citecolor=blue,urlcolor=blue,linkcolor=blue]{hyperref}
\usepackage{amsmath}
\usepackage{amssymb}
\usepackage{babel}
\usepackage{hyperref}
\usepackage{varioref}
\usepackage[active]{srcltx}

\usepackage{bm}
\usepackage{booktabs}
\usepackage{multirow}
\usepackage{array}
\usepackage{enumitem}
\makeatletter

\usepackage{array,multirow,graphicx}
\usepackage{dcolumn}
\usepackage{newlfont}
\usepackage{bm}
\usepackage[colorlinks,citecolor=blue,urlcolor=blue,linkcolor=blue]{hyperref}
\usepackage[figtopcap]{subfigure}
\usepackage{color}

\begin{document}

\date{\today}

\title{Quadratic and cubic spherically symmetric black holes in the modified teleparallel equivalent of general relativity: Energy and thermodynamics}

\author{G.G.L. Nashed}
\email{nashed@bue.edu.eg}
\affiliation {Centre for Theoretical Physics, The British University in Egypt, P.O. Box
43, El Sherouk City, Cairo 11837, Egypt}
\begin{abstract}
In \cite{Bahamonde:2019zea}, a spherically symmetric black hole (BH) was derived from the quadratic form of $f(T)$. Here  we derive the associated
 energy, invariants of curvature, and torsion of this BH and demonstrate  that the higher--order contribution of torsion renders the singularity weaker  compared with the Schwarzschild BH of general relativity (GR). Moreover, we calculate the thermodynamic quantities and reveal  the effect of the  higher--order   contribution on  these quantities. Therefore, we derive a new spherically symmetric BH from the cubic  form of
 $f(T)=T+\epsilon\Big[\frac{1}{2}\alpha T^2+\frac{1}{3}\beta T^3\Big]$, where $\epsilon<<1$, $\alpha$, and $\beta$ are constants. The new BH is
 characterized  by the two constants  $\alpha$ and $\beta$ in addition to $\epsilon$. At $\epsilon=0$ we return to GR. We study the physics of these
 new  BH solutions via the same procedure that was applied for the quadratic BH. Moreover,  we demonstrate  that the contribution of the higher--order torsion, $\frac{1}{2}\alpha T^2+\frac{1}{3}\beta T^3$,  may afford an   interesting physics.
\end{abstract}
\maketitle

\section{Introduction}\label{S1}
The black hole (BH) is the most interesting phenomena in  the general relativity (GR) of Einstein and other modified gravitational theories \cite{Capozziello2011,Clifton_2012,Heisenberg2019,stetsko2020}. This profound and sustained interest in the different approaches to BH physics
can be investigated  because of its relevance in astrophysics \cite{Will2014} and the numerous applications and methods that have initially
 evolved in the gravitational theories  of different systems. Regarding the BH  many topics including the horizons global structure,
 Hawking radiation and thermodynamic properties which are considered as the main goals for realizing the form of the space-time can be studied \cite{Carlip:2019dbu}.
  Although GR is a satisfactory gravitational theory,  modified theories remain greatly desired. It has been  proven that GR  is a
successful theory for isolated masses with length scales of the solar system; however, this theory still faces disputes in the domains of cosmology and quantum scales.

There are many alternative gravitational theories to GR in the literature and the teleparallel
equivalent of general relativity  (TEGR), which was applied by Einstein in (1928) to unify gravity and electromagnetism \cite{2005physics...3046U} is among these theories. In this theory,
the tetrad field is used as a dynamic field instead of the metric of GR. The affine connection is defined regarding the nonsymmetric
 Weitzenb\"ock connection in the TEGR theory whereas the Live-Civita connection plays the affine connection in GR. The use of the Weitzenb\"ock connection affords a space-time without curvature
   and the gravitational field is encoded in the torsion tensor, which is the difference between two Weitzenb\"ock connections.
 The Lagrangian of the TEGR theory depends on the torsion scalar, $T$, which is mainly constructed from the torsion tensor.  There are many physical models in the   solar system regarding the TEGR theory, as  well as in cosmology\cite{andrade2000teleparallel,Nashed:2018urj,aldrov2003selected,Maluf_2013,
 Shirafuji_1996,de_Andrade_1997,El_Hanafy_2016, Nashed_2019,Nashed_2006,Nashed:2009hn,Nashed20062241, NASHED_2007,ULHOA_2010}.

Similar to $f(R)$ which is a generalization of GR in which the Palatini action that depends on the Ricci scalar, $R$, is replaced by an arbitrary
analytic differentiable function \cite{Capozziello_2011,Nojiri_2011,NOJIRI_2007,Nojiri_2017,Nashed:2019tuk}, there is a generalization of the TEGR theory, which is called the  $f(T)$
 gravitational theory \cite{Bengochea:2010sg,Karami_2013,Dent_2011,Cai_2011,Capozziello_2011}. Although TEGR is formulated from a geometry ( Weitzenb\"ock geometry) that is different from GR (the Riemann geometry) both theories are equivalent from the viewpoint of field
 equations. Nevertheless, when we assume the generic forms of  $f(R)$ and $f(T)$, the two theories become inequivalent
\cite{Mai_2017,Ferraro_2008,FIORINI_2009}. The  $f(T)$ theory is an interesting method of  solving the challenges
   of dark energy and dark matters \cite{Cardone_2012,Myrzakulov_2011,Yang_2011,Bamba_2013,Camera_2014,Nashed_2015,Nashed:2013bfa,Wang_2011}. Moreover, in the frame of $f(T)$ there are many interesting BH and cosmological solutions that have  interesting  physics \cite{Rodrigues:2013ifa,Ferraro:2011ks,Nashed:uja,Junior:2015fya,Iorio:2012cm,Xie:2013vua,Ruggiero:2015oka,Awad:2017yod,Awad:2017tyz,Chen:2019ftv,Cai:2015emx}.
  Here we considered the cubic form of the $f(T)$ gravitational theories in the framework of the spherically symmetric space-time to derive a new
  spherically symmetric BH. Additionally, we reconsidered the BH that was derived in \cite{Bahamonde:2019zea} to calculate its associated energy,
   invariants of curvature, and torsion, thereby demonstrating and show how the higher-order contribution of torsion.

  This study is arranged as follows.
 In Section \ref{S11}, we briefly reviewed the TEGR and $f(T)$ formalisms. In  Section \ref{S2}, we discussed the application of a tetrad field that possesses  spherical
 symmetry in four-dimensions  to the vacuum field equations of the $f(T)$ gravity after which the estimated solution  of four-dimensions   is derived for the quadratic form of $f(T)$ \cite{Bahamonde:2019zea}. This BH  asymptotically behaves as the flat space-time. The physical properties of this
  BH  were studied by calculating its invariants in Section \ref{S6336aa}.  In Sections \ref{S6336a} and  \ref{S6336b} we calculated the energy content  and  derived the
  stability condition via the geodesic deviation respectively.  In Section \ref{S3}, we presented  the cubic form of the field equation, using
  $f(T)=T+\epsilon\Big[\frac{1}{2}\alpha T^2+\frac{1}{3}\beta T^3\Big]$,   and derived a new BH solution, up to $O(\epsilon)$,  for these
   differential equations. In Section \ref{S4} the physical properties of the BH which was derived in the cubic form were discussed and analyzed.
     In Sections \ref{S5a} and \ref{S5b},  we calculated the thermodynamic quantities like, Hawking temperature, entropy, heat capacity and Gibbs energy for the quadratic, and  the cubic BH solutions, respectively. The final section was devoted to the discussion and conclusion.

\section{$f(T)$ theory}\label{S11}
 Einstein used  the TEGR theory, which is a gauge theory \cite{2005physics...3046U} to fulfill his dream of unifying gravity with electromagnetism.
In this  theory, $T$  was  responsible  for the gravitational field in the same way $R$ was in the Riemann geometry.
The   affine connection of the TEGR theory  (the Weitzenb\"ock connection)  is defined by the following connection \cite{Wr}
\begin{equation}\label{q2}
 {{W^\lambda}_{\mu \nu}} := {{e_i}^\lambda~ \partial_\nu e^{i}{_{\mu}}},
\end{equation}
where $e^{i}{_{\mu}}$ is the dynamical tetrad in four-dimensions. The metric space-time is defined regardin the tetrad as follows:
\begin{equation}\label{q3}
{g_{\mu \nu} :=  \eta_{i j} {e^i}_\mu {e^j}_\nu,}
\end{equation}
where $\eta_{i j}=(+,-,-,-)$ is a four-dimensional Minkowskian metric of the tangent space.  Using the Weitzenb\" ock connection,
the torsion and contortion tensors are defined as follows:
\begin{eqnarray}
\nonumber {T^\alpha}_{\mu \nu}  & := &
{W^\alpha}_{\nu \mu}-{W^\alpha}_{\mu \nu} ={e_i}^\alpha
\left(\partial_\mu{e^i}_\nu-\partial_\nu{e^i}_\mu\right),\\
{\gamma^{\mu \nu}}_\alpha  & := &
-\frac{1}{2}\left({T^{\mu \nu}}_\alpha-{T^{\nu
\mu}}_\alpha-{T_\alpha}^{\mu \nu}\right). \label{q4}
\end{eqnarray}
It is well known that the difference between Weitzenb\"ock and Levi-Civita connections reproduces the contortion, as follows:
\begin{equation}
{\gamma^\mu}_{\nu \alpha}={W^\mu}_{\nu \alpha}-{\Gamma^\mu}_{\nu \alpha}.\end{equation}
The superpotential tensor ${\Sigma_\mu}^{\nu \alpha}$ which is the antisymmetric tensor in the last two indices is defined as follows:
\begin{equation}\label{q5}
{\Sigma_\alpha}^{\mu \nu} := \frac{1}{2}\left({\gamma^{\mu\nu}}_\alpha+\delta^\mu_\alpha{T^{\beta
\nu}}_\beta-\delta^\nu_\alpha{T^{\beta \mu}}_\beta\right).
\end{equation}
Using all the above data we can define $T$ of the TEGR theory according to Eq. (\ref{Tor_sc}):
\begin{equation}\label{Tor_sc}
T := {T^\alpha}_{\mu \nu} {\Sigma_\alpha}^{\mu \nu},
\end{equation}

Similar to the  GR  extension  ($f(R)$)  it was logical to modify the TEGR theory to include higher torsion
orders, which enabled us to define the  Lagrangian of $f(T)$, where $f$ is an analytical continues differentiable function of $T$:
\begin{equation}\label{q7}
{\cal L}=\frac{1}{2\kappa}\int |e|f(T)~d^{4}x,
\end{equation}
where $|e|=\sqrt{-g}=\det\left({e^a}_\mu\right)$ is the determinate of the metric and  $\kappa$
is  a four-dimensional constant that is defined as $\kappa =\frac{8\pi G}{c^4}$, where  $G$ is the Newtonian gravitational  constant in
four-dimensions and  $c$ is the speed of light.  The variation of Eq. (\ref{q7}) w.r.t.  the  tetrad field, ${e^i}_\mu$
 affords the following  field equations of $f(T)$ in the vacuum case as shown in \cite{Bengochea:2010sg}:
\begin{eqnarray}\label{q8}
{S_\mu}^{\rho \nu} \partial_{\rho} T f_{TT}+\left[e^{-1}{e^i}_\mu\partial_\rho\left(e{e_i}^\alpha
{S_\alpha}^{\rho \nu}\right)-{T^\alpha}_{\lambda \mu}{S_\alpha}^{\nu \lambda}\right]f_T
-\frac{f}{4}\delta^\nu_\mu =\mathfrak{\xi}^\nu{}_\mu \equiv0,
\end{eqnarray}
where $f := f(T)$, \ \   $f_{T}:=\frac{\partial f(T)}{\partial T}$ and  $f_{TT}:=\frac{\partial^2 f(T)}{\partial T^2}$.
The application of the field equations (\ref{q8}) to a spherically symmetric tetrad field using the form of
$f(T)=T+\epsilon\Big[\frac{1}{2}\alpha T^2+\frac{1}{3}\beta T^3\Big]$ is shown in Section \ref{S3}.

\section{Asymptotically stationary  AdS black holes}\label{S2}
In this section, the  field equations of the higher order torsion theory, (Eq. (\ref{q8})) were applied  to the spherically symmetric space-time,
 thus affording the vielbein  which is written in the spherical  coordinate ($t$, $r$, $\theta$, $\phi$) as shown in~\cite{Bahamonde:2019zea}:
\begin{equation}
e^a{}_{\nu}=\left(
\begin{array}{cccc}
\sqrt{\mu} & 0 & 0 & 0 \\
0 & \sqrt{\nu} \cos (\phi ) \sin (\theta ) & r \cos (\phi ) \cos (\theta )  & -r \sin (\phi ) \sin (\theta )  \\
0 & \sqrt{\nu} \sin (\phi ) \sin (\theta )  & r \sin (\phi ) \cos (\theta )  & r \cos (\phi ) \sin (\theta ) \\
0 & \sqrt{\nu} \cos (\theta ) & -r \sin (\theta ) & 0 \\
\end{array}
\right)\label{tetrad}\,.
\end{equation}
where $-\infty < t < \infty$,  $0\leq r< \infty$, $0\leq \theta< \pi$,  $0\leq \phi< 2\pi$, $\mu$ and $\nu$ are the two unknown functions of $r$.  Thus, the space-time, which can be generated by (\ref{tetrad}) is expressed as follows:
 \begin{eqnarray} \label{m2}
ds^2=\mu(r) \,dt^2-\nu(r)\, dr^2-r^2d\Sigma^2\,, \end{eqnarray}
where $d\Sigma^2=d\theta^2+\sin^2\theta d\phi^2$ is the two dimensional sphere.
Substituting  Eq. (\ref{tetrad}) into Eq. (\ref{Tor_sc}), we evaluate $T$ as follows\footnote{The abbreviations are represented as follows $\mu(r)\equiv \mu$,
 \ \ $\nu(r)\equiv \nu$, \ \ $\mu'\equiv\frac{d\mu}{dr}$ and $\nu'\equiv\frac{d\nu}{dr}$ .}
\begin{equation}\label{df}
T=  -\frac{2 \left(\sqrt{\nu(r)}-1\right) \left(r \mu'(r)-\mu(r) \sqrt{\nu(r)}+\mu(r)\right)}{r^2 \mu(r) \nu(r)}\,..
\end{equation}
Applying Eq. (\ref{tetrad}) to the vacuum field equation (\ref{q8}) the following nonvanishing components could be obtained \cite{Bahamonde:2019zea}:
\begin{eqnarray}
\xi^t{}_t&=&\frac{r \beta(\sqrt{\nu}-1) \mu'+\mu (r \nu'+2 \nu^{3/2}-2 \nu)}{2 r^2 \mu \nu^2}f_T+\frac{(\sqrt{\nu}-1) }{r \nu}T' f_{TT}+\frac{1}{4} f \,,\label{Eq1}\\
\xi^r{}_r&=& - \frac{r (\sqrt{\nu}-2) \mu'+2 \mu (\sqrt{\nu}-1) }{2 r^2 \mu \nu}f_{T} - \frac{1}{4} f\,,\label{Eq2}\\
\xi^\theta{}_\theta=\xi^\phi{}_\phi&=&\frac{-r^2 \nu \mu'^2+r \nu \left(-r \mu' \nu'-4 \nu^{3/2} \mu'+2 \nu\left(r \mu''+3 \mu'\right)\right)+\mu^2
\left(-2 r \nu'-8 \nu^{3/2}+4 \nu^2+4 \nu\right)}{8 r^2 \mu^2 \nu^2}f_{T}\nonumber\\
&&+\frac{r \mu'-2 \mu (\sqrt{\nu}-1) }{4 r \mu(r) \nu(r)}T'f_{TT}-\frac{1}{4} f\,,\label{Eq3b}
\end{eqnarray}
Bahamonde et al. \cite{Bahamonde:2019zea} have solved the above system when $f(T)=T+\frac{1}{2}\epsilon\,\alpha T^p$.
In this section, we  discussed the physics of the BH solution that was derived in \cite{Bahamonde:2019zea}  for the case ``p=2'':\\
\begin{equation}\label{bfin}
\mu(r)=\sigma+\epsilon\,\sigma_1\,, \qquad \qquad \nu(r)=\sigma^{-1}+\epsilon\, \sigma_2\,,
\end{equation}
where
\begin{eqnarray}
 \sigma &=& 1- \frac{2m}{r}\,,\nonumber\\
  \sigma_1&=&\left[-\frac{c_1}{r}+c_2-\alpha\left(\frac{m^2+6 m r+r^2}{m r^3}-\frac{16 \left(1-\frac{2 m}{r}\right)^{3/2}}
    {3 m^2}+\frac{(1-\frac{3 m}{r})}{2 m^2} \ln \left(1-\frac{2m}{r}\right)\right)\right]\,,\nonumber\\
 \sigma_2&=&\left[\frac{\left(\frac{c_1}{r}-\frac{2 c_2 m}{r}\right)}{\left(1-\frac{2 m}{r}\right)^{2}}
     -\alpha\left(-\frac{8 (3 m^2-7 m r+2 r^2)}{3 m r^3 (1-\frac{2 m}{r})^{3/2}} +\frac{25  m-23 r}{r^3 (1-\frac{2 m}{r})^2}+
     \frac{\ln\left(1-\frac{2m}{r}\right)}{2mr(1-\frac{2m}{r})^{2}}\right)\right]\,,
\end{eqnarray}
with $c_1$ and $c_2$ are the constants of the integration. Equation (\ref{bfin}) could be reduced to the Schwarzschild BH GR when $\epsilon=0$. In the
following subsections we  extracted the physics of Eq. (\ref{bfin}) when $\epsilon \neq 0$. From now on  we will refer to solution (\ref{bfin}) as ``p=2''.
\subsection{Singularities of ``p=2''}\label{S6336aa}
We calculate the invariant of Eq. (\ref{bfin}), up to $O(\epsilon)$, and obtained the following
\begin{eqnarray}\label{inv}  &&  T^{\mu \nu \lambda}T_{\mu \nu \lambda} = -\frac{16\epsilon\, m}{r^2}+\frac{2m^2}{r^4}\nonumber\\
 &&
+\epsilon\Bigg[\frac{8c_1+12\alpha ln(2)/m+128\alpha/3m-16mc_2}{r^3}+
  \frac{2mc_1+3\alpha ln(2)+32\alpha/3-4m^2c_2}{r^4}\Bigg]+O\Big(\frac{1}{r^6}\Big),\nonumber\\
 &&T^\mu T_\mu = -\frac{16\epsilon\, m}{r^2}-\frac{m^2}{r^4}-\frac{4m^3}{r^5}+\epsilon\Bigg[\frac{64\alpha/3m-8mc_2+6\alpha ln(2)/m+4c_1}{r^3}\nonumber\\
 &&
+ \frac{2m^2c_2-16\alpha/3-3\alpha ln(2)/2-mc_1}{r^4}+\frac{12m^3c_2-32m\alpha-9m\alpha ln(2)-6m^2c_1}{r^5}\Bigg]+O\Big(\frac{1}{r^6}\Big), \nonumber\\
 &&T(r)=\frac{8\epsilon\,m}{r^2}+\frac{2m^2}{r^4}+\frac{4m^3}{r^5}+\epsilon\Bigg[\frac{3\alpha\, ln(2)+2m\,c_1+32\alpha/3-4m^2\,c_2}{r^4}+
  \frac{9m\,\alpha\, ln(2)+6\,m^2\,c_1+32\,m\,\alpha-12\,m^3\,c_2}{r^5}\Bigg]+O\Big(\frac{1}{r^6}\Big), \nonumber\\
&& R^{\mu \nu \lambda \rho}R_{\mu \nu \lambda \rho}=-\frac{48m^2}{r^6}+\epsilon\Bigg[\frac{96m^2c_2-48mc_1-256\alpha-72\alpha ln(2)}{r^6}\Bigg]
+O\Big(\frac{1}{r^{10}}\Big),
\nonumber\\
&&
R^{\mu \nu}R_{\mu \nu}= O\Big(\epsilon^2 \Big)\,,
 \qquad R= \epsilon\Bigg[\frac{16m^3\alpha}{r^7}+\frac{20m^4\alpha}{r^8}\Bigg]+O\Big(\frac{1}{r^{9}}\Big),
  \end{eqnarray}
where  $T^{\mu \nu \lambda}T_{\mu \nu \lambda}$ $T^{\mu }T_{\mu }$, $T$, $R^{\mu \nu \lambda \rho}R_{\mu \nu \lambda \rho}$, $R^{\mu \nu}R_{\mu \nu }$ and
  $R$, are the torsion tensor square,  torsion vector square,  torsion scalar,  Kretschmann scalar,  Ricci tensor square, and Ricci scalar.  The above invariants indicated that there was a singularity (curvature singularity) at $r=0$.  Close to $r=0$,  the behaviors  of  $T^{\mu \nu \lambda}T_{\mu \nu \lambda}$, $T^{\mu }T_{\mu }$, $T$  are given by $T^{\alpha \beta \gamma}T_{\alpha \beta \gamma}=T^{\alpha}T_{\alpha}=T\sim  {r^{-2}}$
  in contrast with the  solutions of the Einstein theory in  GR and TEGR, which are given  as $T^{\alpha \beta \gamma}T_{\alpha \beta \gamma}=T^{\alpha}
  T_{\alpha}=T\sim  {r^{-4}}$. This demonstrates that the singularity of the higher-order torsion theory was much milder than the one obtained in GR
  and TEGR for the natural spherically symmetric case. This result suggests that these singularities are {\textrm weak ones},
  according to Tipler and Krolak \cite{CLARKE1985127,TIPLER19778}, and the  possibility of extending the geodesics beyond these regions.
   This would be discussed in subsequent  studies.
\subsection{Energy content of ``p=2''}\label{S6336a}
Here, we  calculated the energy content of Eq. (\ref{bfin}). To do this, we applied the Hamiltonian density  $H$ which
 can be obtained from the Lagrangian equation $(\ref{q7})$ by  rewriting
it as follows: \[{\cal L} = p\dot{q}-H.\]
 The Hamiltonian of the $f(T)$ gravitational theory is expressed as follows:
 \begin{eqnarray} \label{h1}
 H=e^a{}_{0} B^a+\xi_{ij}\Gamma^{ij}+\xi_{j}\Gamma^{j}, \qquad \xi_{ij}=\frac{1}{2}e_i{}^\alpha e_j{}^\beta (T_{\alpha 0 \beta} - T_{\beta 0 \alpha}),
 \qquad \xi_{j}=e_0{}^\alpha e_j{}^\beta T_{\alpha 0\beta},\end{eqnarray} where $\xi_{ij}$ and $\xi_j$  are the Lagrange multipliers.
The components  $e^a{}_{0}$  do not exhibit time dependence thus this quantity could be considered as a
Lagrange multiplier. The canonically conjugate moneta  to $e^a{}_{\alpha}$ are denoted
by $\Pi^{a\alpha}$\footnote{Notably,  the Latin indices were raised and lowered by the Minkowski metric.}. In the configuration space one can obtain  \cite{ULHOA_2013}:
\begin{eqnarray}
\Pi^{a c} =e^a{}_\alpha e^c{}_\beta \Pi^{\alpha \beta} = -4\pi e^a{}_\alpha e^c{}_\beta \Sigma^{\alpha 0\beta}f_T= -4\pi  \Sigma^{a 0 c}f_T .\end{eqnarray}
Therefore,  writing $H$  in terms of $e^a_{\alpha}$, $\Pi^{a\alpha}$ and Lagrange
multipliers is possible. To express a simple  form of  $H$    some Lagrange multipliers and constraints
 were redefined \cite{Maluf_2006} and  Eq. (\ref{h1}) reads as follows:
\begin{eqnarray}
H = e_{a 0}C^a+\frac{1}{2}\lambda_{ab}\Gamma^{ab}, \end{eqnarray}
where $e_{a0}$ and $\lambda_{ab} = -\lambda_{ba}$ are the Lagrange multipliers, and  $C^a$ and $\Gamma^{ab}$ are the first-class constraints. Solving the Hamilton field equations can aid the
identifications of  $\lambda_{ij}=e_i{}^\alpha e_j{}^\beta \lambda_{\alpha \beta} = \frac{1}{2}e_i{}^\alpha e_j{}^\beta (T_{\alpha 0 \beta} -
T_{\beta 0 \alpha})$ and $\lambda_{0j}=e_0{}^\alpha e_j{}^\beta \lambda_{\alpha \beta} =e_0{}^\alpha e_j{}^\beta T_{\alpha 0 \beta}$.
 The quantities $\lambda_{ij}$ and $\lambda_{0j}$
are the components of Eq. (\ref{Ca}) \begin{eqnarray}\label{Ca} \lambda_{\mu \nu} =e^a{}_\mu e^b{}_\nu \lambda_{ab}.\end{eqnarray}
The constraint $C^a$ may be written in the following form: \begin{eqnarray} C^a = -\partial_i \Pi^{ai}f_T + f^a,\end{eqnarray} where
$f^a$ is a lengthy  expression of the field quantities. Notably $\partial_i \Pi^{ai}$ is
the only total divergence term of the momenta $\Pi^{ai}$, which emerged  in the expression of
$C^a$. The constraint $C^a = 0$ inspired the definition of the
gravitational energy-momentum $P^a$  in four-dimensions in the integral form:
\begin{eqnarray} \label{en}
P^a=-\int_V d^3x \partial_i\Pi^{ai},\end{eqnarray}
where  $V$ is the three-dimensional volume of the  space \cite{ULHOA_2013}.

  Next the  energy that was related to  the  BH was calculated using Eq. (\ref{bfin}). Using Eq. (\ref{en}),
   the necessary components   for calculating the energy
    in the following  form could be derived\footnote{The square parentheses in the quantities  $\Sigma^{(0)(0)1}$  refer to the tangent components, i.e.,
     $\Sigma^{(0)(0)1=e^0{}_\mu e^0{}_\nu \Sigma^{\mu \nu 1}}$.}:
\begin{eqnarray} \label{se}  \Sigma^{(0)(0)1}=\frac{r\sin(\theta)\sqrt{\mu \nu}(\sqrt{\nu}-1)}{\nu}.  \end{eqnarray}
 Substituting  Eqs. (\ref{bfin}) and (\ref{se}) into Eq. (\ref{en}),  we obtained the energy content of the black hole (\ref{bfin}), up to $O(\epsilon)$, in the following form:
\begin{eqnarray}  \label{en1} &&P^0=E\approx m+\epsilon\frac{44\alpha+24mc_1+3\alpha ln(2)-4m^2c_2}{8m}+O\left(\frac{1}{r}\right), \nonumber\\
  && \end{eqnarray}
which obtained  $E=m$ when $\epsilon\rightarrow 0$ which is ADM (Arnowitt, Deser Misner) mass \cite{Misner:1974qy}. Equation (\ref{en1}) is  finite value and indicates that the energy  depended on
 the coefficient of the higher-order torsion terms, $\epsilon$, up to order O$\left(\frac{1}{r}\right)$. Moreover,  Eq. (\ref{en})
 indicates that the values of $\epsilon$, $\alpha$ and $c_1$ must be positive and that of $c_2$ must be negative value or $8m^2+
 \epsilon(44\alpha+24mc_1+3\alpha ln(2))>4\epsilon m^2c_2$.
\subsection{Analysis of the stability of the  black hole   with the geodesic deviation}\label{S6336b}
 The paths  of a test particle in the gravitational field are described by the following equation:
 \begin{equation}\label{ge}
 {d^2 x^\alpha \over d\tau^2}+ \Bigl\{^\alpha_{ \beta \rho} \Bigr \}
 {d x^\beta \over d\tau}{d x^\rho \over d\tau}=0,
 \end{equation}
which is known as the geodesic equations.  In Eq. (\ref{ge})  $\tau$ represents  the affine  connection parameter. The
  geodesic  deviation  possesses the form \cite{1992ier..book.....D,Nashed:2003ee}
  \begin{equation} \label{ged}
 {d^2 \xi^\sigma \over d\tau^2}+ 2\Bigl\{^\sigma_{ \mu \nu} \Bigr \}
 {d x^\mu \over d\tau}{d \xi^\nu \over d\tau}+
\Bigl\{^\sigma_{ \mu \nu} \Bigr \}_{,\ \rho}
 {d x^\mu \over d\tau}{d x^\nu \over d\tau}\xi^\rho=0,
 \end{equation}
where $\xi^\rho$ is the four-vector deviation. Introducing (\ref{bfin}) into (\ref{ge}) and (\ref{ged}), obtained the following:
\begin{equation}
{d^2 t \over d\tau^2}=0, \qquad {1 \over 2} \mu'(r)\left({d t \over
d\tau}\right)^2-r\left({d \phi \over d\tau}\right)^2=0, \qquad {d^2
\theta \over d\tau^2}=0,\qquad {d^2 \phi \over d\tau^2}=0,\end{equation} and for
the BH regarding the geodesic deviation  (\ref{bfin}) afforded the following:  \begin{eqnarray}\label{ged11} && {d^2 \xi^1 \over d\tau^2}+\nu(r)\mu'(r) {dt \over d\tau}{d
\xi^0 \over d\tau}-2r \nu(r) {d \phi \over d\tau}{d \xi^3 \over
d\tau}+\left[{1 \over 2}\left(\mu'(r)\nu'(r)+\nu(r) \mu''(r)
\right)\left({dt \over d\tau}\right)^2-\left(\nu(r)+r\nu'(r)
\right) \left({d\phi \over d\tau}\right)^2 \right]\xi^1=0, \nonumber\\
&&  {d^2 \xi^0 \over
d\tau^2}+{\nu'(r) \over \nu(r)}{dt \over d\tau}{d \zeta^1 \over d\tau}=0,\qquad {d^2 \xi^2 \over d\tau^2}+\left({d\phi \over d\tau}\right)^2
\xi^2=0, \qquad \qquad  {d^2 \xi^3 \over d\tau^2}+{2 \over r}{d\phi \over d\tau} {d
\xi^1 \over d\tau}=0, \end{eqnarray} where $\mu(r)$ and $\nu(r)$ were defined from Eq.  (\ref{bfin}),
$\nu'(r)=\displaystyle{d\alpha(r) \over dr}$. Using
the circular orbit
\begin{equation} \theta={\pi \over 2}, \qquad
{d\theta \over d\tau}=0, \qquad {d r \over d\tau}=0,
\end{equation}
we get
\begin{equation}
 \left({d\phi \over d\tau}\right)^2={\mu'(r)
\over r[2\mu(r)-r\mu'(r)]}, \qquad \left({dt \over
d\tau}\right)^2={2 \over 2\mu(r)-r\mu'(r)}. \end{equation}

\begin{figure}
\centering
\subfigure[~Stability of (\ref{stab1})]{\label{fig:1}\includegraphics[scale=0.3]{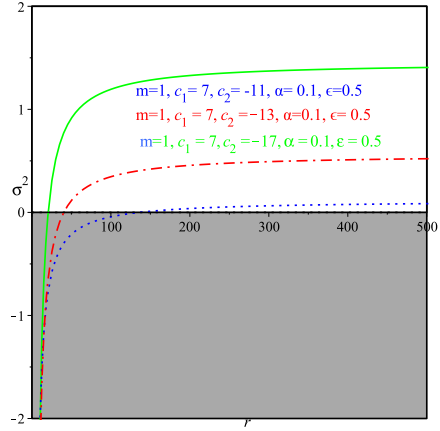}}\hspace{0.2cm}
\caption{ {Schematic plot of Eq. (\ref{stab1}), namely, $\sigma^2$  versus the coordinate, $r$.}}
\label{Fig:1}
\end{figure}

Further, Eqs.~(\ref{ged11}) can be rewritten as follows:
\begin{eqnarray} \label{ged22} &&  {d^2 \xi^1 \over d\phi^2}+\mu(r)\mu'(r) {dt \over
d\phi}{d \xi^0 \over d\phi}-2r \mu(r) {d \xi^3 \over
d\phi} +\left[{1 \over 2}\left[\mu'^2(r)+\mu(r) \mu''(r)
\right]\left({dt \over d\phi}\right)^2-\left[\mu(r)+r\mu'(r)
\right]  \right]\zeta^1=0, \nonumber\\
&&{d^2 \xi^2 \over d\phi^2}+\xi^2=0, \qquad {d^2 \xi^0 \over d\phi^2}+{\mu'(r) \over
\mu(r)}{dt \over d\phi}{d \xi^1 \over d\phi}=0,\qquad {d^2 \xi^3 \over d\phi^2}+{2 \over r} {d \xi^1 \over
d\phi}=0. \end{eqnarray}
The second equation  of Eq. (\ref{ged22}) corresponds to a simple harmonic motion, which  indicates the  stability on the plane $\theta=\pi/2$,
assuming  the remaining equations of (\ref{ged22}) obtained solutions in the form of Eq. (\ref{ged33}):
\begin{equation} \label{ged33}
\xi^0 = \zeta_1 e^{i \sigma \varphi}, \qquad \xi^1= \zeta_2e^{i \sigma
\varphi}, \qquad and \qquad \xi^3 = \zeta_3 e^{i \sigma \varphi},
\end{equation}
where $\zeta_1, \zeta_2$ and $\zeta_3$ are the constants and  $\varphi$ is an unknown variable. Substituting  Eq. (\ref{ged33}) into
(\ref{ged22}),  the stability condition for  static spherically symmetric charged BH can be obtained in the following  form:
\begin{equation} \label{con1}
 \frac{3\mu\nu\nu'-\sigma^2\mu \nu'-2r\nu^{3/2}\mu'^{3/2}-r\mu\nu'^2+r\mu\nu'\mu'+r\mu\nu \mu''}{\mu \nu'}>0.
\end{equation}
 Equation (\ref{con1}) obtained the following solution:
\begin{equation} \label{stab1}
\sigma^2= \frac{3\mu \nu\nu'-2r\nu^{3/2}\mu'^{3/2}-r\mu\nu'^2+r\mu\nu'\mu'+r\mu\nu \mu''}{\mu^2 \nu'^2}>0.\end{equation}
Figure \ref{Fig:1} which  exhibits the regions where the BH solutions are stable and the regions where there are no possible stability is a plot of Eq. (\ref{stab1}) for particular values of the model.
\section{Cubic solution of Eq. (\ref{q8})}\label{S3}
In this section, we derived a novel spherically symmetric solution using the form $f(T)=T+\epsilon\,[\frac{1}{2}\alpha T^2+\frac{1}{3}\beta T^3]$.
 To do this, we are derived the cubic form of the field equations (\ref{q8}) as follows:
\begin{eqnarray}
&&\xi^t{}_t=\frac{1}{3r^6\varrho^2}\Bigg[3r^4\varrho^2(1-\varrho-r\varrho')+\epsilon\Big\{12r^2\varrho''\Big[\varrho^{5/2}(16\beta-12r\beta\varrho'+2r^2\alpha)
+4\varrho^{3/2}r\beta\varrho'+16\varrho^{7/2}\beta-\varrho^2(4r\beta\varrho'[3+\varrho]\nonumber\\
&&+4\beta \varrho^2+\varrho[24\beta+r^2\alpha]+4\beta+r^2\alpha)\Big]+6r\varrho'\Big[6\varrho^{5/2}(r^2\alpha-4\beta)-24\varrho^{7/2}\beta+\varrho^{2}
(6\beta-3r^2\alpha+\varrho\{r^4\mu_1-3r^2\alpha\nonumber\\
&&+36\beta\}+6\varrho^2\beta)\Big]+3r^2\varrho'^2[r^2\alpha\varrho-48\beta\varrho^2-3r^2\alpha \varrho^2+4\beta \varrho-2[4\beta+r^2\alpha +28\varrho\beta]\varrho^{3/2} -20\beta\varrho^3]
+3 r^3\beta\varrho^{1/2}\varrho'^3\Big[3\varrho^{1/2}-4\nonumber\\
&&+6\varrho-5\varrho^{3/2}\Big]+\varrho^2\Big[3r^2\alpha-24[4\beta+r^2\alpha]\varrho^{1/2}+6\varrho(70\beta+9r^2\alpha)-8\varrho^{3/2}(6r^2\alpha+100\beta)+3\varrho^2
\Big(260\beta+r^5\nu'_1\nonumber\\
&&+\nu_1 r^4+5r^2\alpha\Big)+\beta\varrho^{5/2}(76\varrho^{1/2}-384)\Big]\Big\}\Bigg] \,,\nonumber\\
&&\xi^r{}_r=\frac{1}{3r^6\varrho^{3/2}}\Bigg[3r^4\varrho^{3/2}(1-\varrho-r\varrho')+\epsilon\Big\{3r\varrho'\Big[\varrho^{5/2}\{r^4\nu_1 -6(r^2\alpha+20\beta)\}
-2\varrho^{3/2}(10\beta+3r^2\alpha)-20\varrho^{7/2}\beta+r^4\mu_1\varrho^{1/2}\nonumber\\
&&+80\beta\varrho^3+4\varrho^2\{20\beta+3r^2\alpha\}\Big]-3r^2\varrho'^2\Big[(72\beta+20\varrho+3r^2\alpha) \varrho^{3/2}-\varrho^{1/2}(32\varrho^{1/2}\beta+
4r^2\alpha \varrho^{1/2}+4\beta+r^2\alpha-64\beta \varrho^{3/2})\Big]\nonumber\\
&&+4r^3\beta\varrho'^3\Big[2+12\varrho-\varrho^{1/2}(9+5\varrho)\Big]+\varrho^{3/2}\Big[3r^2\alpha+4\beta-3r^5\mu'_1-6\varrho(3r^2\alpha-10\beta)+8\varrho^{3/2}
(3r^2\alpha+20\beta)-3\varrho^2\Big(60\beta+3r^2\alpha\nonumber\\
&&-r^4\nu_1\Big)-4\beta\varrho^{5/2}(24-5\varrho^{1/2})\Big]\Big
\}\Bigg]\,,\nonumber\\
&&\xi^\theta{}_\theta=\xi^\phi{}_\phi=
\frac{1}{12r^6\varrho^2}\Bigg[6r^5\varrho^2(2\varrho'+r\varrho'')-\epsilon\Big\{6r^2\varrho''\Big[4r\varrho^{3/2}\varrho'(r^2\alpha+8\beta+6r\beta\varrho')+
4\varrho^{5/2}(20\beta+3r^2\alpha+24r\beta\varrho')+80\varrho^{7/2}\beta\nonumber\\
&&+\varrho\Big(r^4\mu_1-2\varrho\{3r^2\alpha-10\beta\}+\varrho^2\{r^4\nu_1-6r^2\alpha-120\beta\}
-20\beta \varrho^3-4r\varrho\varrho'\{r^2\alpha+24\beta+8\beta \varrho\}-12r^2\beta\varrho'^2\{1+\varrho\}\Big)\Big]\nonumber\\
&&+3r\varrho'\Big[8\varrho^{3/2}(r^2\alpha+4\beta)+
24r^2\varrho^{5/2}\alpha-160\beta\varrho^{7/2}+\varrho\Big(2r^4\mu_1-8\varrho\{3r^2\alpha-10\beta\}+2\varrho^2\{3r^4\nu_1-4r^2\alpha+80\beta\}\nonumber\\
&&
+48\beta \varrho^3+r^5\varrho^2\nu'_1+r^5\mu'_1\Big)\Big]+3r\varrho'^2\Big[4\varrho^{3/2}(3r^2\alpha+20\beta)+
112\beta\varrho^{5/2}+32\beta\varrho^3+\varrho^2(12r^2\alpha-r^4\nu_1+144\beta)+16\varrho\beta+r^4\mu_1\Big]
\nonumber\\
&&
+2r^3\varrho^{1/2}\varrho'^3\Big[60\varrho\beta-3r^2\alpha-40\varrho^{3/2}\beta-20\beta\Big]+24r^4\beta\varrho'^4\Big[1-\varrho^{1/2}\Big]-
2\varrho^2\Big[6r^2\alpha+16\beta-12\varrho(3r^2\alpha+20\beta)\nonumber\\
&&+16\varrho^{3/2}(40\beta+3r^2\alpha)-\varrho^2(720\beta+3r^5\nu'_1+18r^2\alpha)
-16\beta\varrho^{5/2}(5\varrho^{1/2}-24)\Big]\Big\}\Bigg]\,,\label{Eq3}
\end{eqnarray}
where $\varrho=1-\frac{2M}{r}$ and we have used
\begin{eqnarray} \label{pot} \mu=\varrho+\epsilon\mu_1, \qquad  \textrm {and} \qquad   \nu=\varrho^{-1}+\epsilon\nu_1.\end{eqnarray}
 From the second equation of Eq. (\ref{Eq3}) we obtained the following expression:
\begin{eqnarray} \label{Eq33}
&&\nu_1=\frac{1}{3r^{7/2}\varrho^{5/2}}\Big[\varrho^{1/2}[3r^5\varrho\mu'_1+3\varrho^2(\alpha r^2+16\beta)+\varrho(3\mu_1 r^4+18\alpha r^2+160\beta)
+48\beta+3\alpha r^2-3\mu_1 r^4]-4(1+\varrho)(2\beta \varrho^2\nonumber\\
&&+\varrho[3\alpha r^2+28\beta]+2\beta)\Big]\,.\label{sol1}
\end{eqnarray}
 Further, using Eq. (\ref{Eq33}) in the third equation of  (\ref{Eq3}) we obtained the following expression:
 \begin{eqnarray}
&&\mu_1=\frac{1}{1890r^4 \varrho(1-\varrho)^4}\Big[10r^2\varrho^4\{189r^2c_4\varrho+5859\alpha\}+10\beta\varrho^{1/2}[1008-4032\varrho
-112\varrho^{6}+34608\varrho^2+448\varrho^{5}]\nonumber\\
&&-10r^2\alpha\varrho[945\varrho ln(\varrho)-1323\varrho^2 ln(\varrho)+567\varrho^3 ln(\varrho)-
189 ln(\varrho)+567 ln(2)-1134 ln(2)\varrho+567ln(2)\varrho^2]\nonumber\\
&&-\beta \varrho\Big[105840ln(2)+105840ln(\varrho)\varrho-
15120 ln(\varrho)+150528\varrho^{5/2}\Big]+10c_4r^4\varrho[189-756\varrho+1134\varrho^2-756\varrho^3]\nonumber\\
&&+10c_3r^3\varrho[756r\varrho^3-189+756\varrho-1134\varrho^2-
189\varrho^4]-16065r^2\alpha r^2\varrho-10\beta\varrho[3318\varrho^4+2562+2226\varrho-504\varrho^2\nonumber\\
&&-7224\varrho^3]
+\alpha\varrho^2 r^2[61425-88830\varrho-16065\varrho^3+945\varrho^4]+3780\beta\varrho^6\nonumber\\
&&+27936\beta\varrho^{9/2}+10r^2\alpha\varrho^{5/2}\{4032\varrho^{2}-8064\varrho+4032\}\Big]\,.\label{sol2}
\end{eqnarray}
Substituting  Eq. (\ref{sol2}) in Eq. (\ref{sol1}) we get the following:
 \begin{eqnarray}
&&\nu_1=\frac{1}{1890r^4 \varrho^{5/2}(1-\varrho)^4}\Big[9450c_4r^4\varrho^{3/2}\{1-\varrho^3\}+10c_3r^2\varrho^{3/2}[189r\varrho^3-756r\varrho^2
-756r]+10\varrho^{3/2}[31752\varrho \beta ln(\varrho)-4158r^2\alpha]\nonumber\\
&&+r^2\alpha\varrho^{1/2}[5670\varrho^3 ln(2)-3780\varrho^2 ln(\varrho)-5670\varrho^2 ln(2)+17955]-10\beta \varrho^{1/2}[1512ln(\varrho)-10584ln(2)]+
1890r^3\varrho^{1/2}[c_3-rc_4]\nonumber\\
&&+10\alpha r^2\varrho^{1/2}[2520\varrho^{7/2}+567 ln(2)-189ln(\varrho)]-10\beta[1512-5586\varrho^{1/2}-7896\varrho^{3/2}-
1890\varrho^{13/2}-12978\varrho^{5/2}+37506\varrho^{9/2}\nonumber\\
&&-7560\varrho^{11/2}-27552\varrho^{7/2}]+11340r^3\varrho^{5/2}c_3+18900r^4 c_4\varrho^{5/2}[\varrho-1]+r^2\alpha\varrho^{1/2}
[18900\varrho^5+5670\varrho^4 ln(\varrho)+7560\varrho ln(\varrho)\nonumber\\
&&-5670\varrho ln(2)+945\varrho^6-7560\varrho^3
ln(\varrho)-7560\varrho^{1/2}]+\beta\varrho[108272\varrho^4-45360-28560\varrho-1023792\varrho^2+287568\varrho^3]\nonumber\\
&&+r^2\alpha\varrho^{2}
\{25200\varrho-17640-17640\varrho^3-7560\varrho^4-74655\varrho^{5/2}+83160\varrho^{3/2}-4725\varrho^{1/2}\}+10\beta\varrho^{3/2}\{6048ln(\varrho)+
31752ln(2)\nonumber\\
&&-6776\varrho^{9/2}-168\varrho^{11/2}\}+1890r^4c_4\varrho^{11/2}\Big]\,.\label{sol3}
\end{eqnarray}
From  Eqs. (\ref{sol2}) and (\ref{sol3}) we constructed the unknown functions $\mu$ and $\nu$ in the asymptotic form up to $O(\epsilon)$. Notably,  the metric potential (\ref{pot}) using Eq. (\ref{sol2}) and (\ref{sol3}) was the solution to the field equations (\ref{q8})
  up to $O(\epsilon)$ for the form $f(T)=T+\epsilon\Big[\frac{1}{2}\alpha T^2+\frac{1}{3}\beta T^3\Big]$. In Section \ref{S4} we extracted the physics of the solution of (\ref{pot}) using Eqs. (\ref{sol2}) and (\ref{sol3}).
\section{Main features of the cubic solution}\label{S4}
Some features of the solution that was derived in the previous section  were analyzed here.\\
\underline{The asymptote of the metric:}\\
By constructing the metric of solution (\ref{pot}) from Eqs. (\ref{sol2}) and (\ref{sol3}) we could easily demonstrate  that this solution asymptotically behaved as a
flat space-time and that is an acceptable behavior.
\subsection{Singularities of the cubic solution }\label{S6336}
We calculated the invariant of  solution (\ref{pot}) from Eqs. (\ref{sol2}) and (\ref{sol3}) and obtained the following:
\begin{eqnarray}\label{invc}  &&  T^{\mu \nu \lambda}T_{\mu \nu \lambda} = -\frac{16\epsilon\, M}{r^2}+
\frac{-45360M^4+\epsilon[M^3(15120c_3-30240Mc_4)+\alpha M^2(80640+22680ln(2))+\beta(105840ln(2)+32768)]}{7560M^2 r^4}\nonumber\\
 &&
+O\Big(\frac{1}{r^5}\Big),\nonumber\\
 &&T^\mu T_\mu = -\frac{16\epsilon\, M}{r^2}+\frac{-5040M^4+\epsilon[M^3(15120c_3-30240Mc_4)+\alpha M^2(80640+22680ln(2))+\beta(105840ln(2)+32768)]}{5040M^2 r^4}\nonumber\\
 &&
+O\Big(\frac{1}{r^5}\Big), \nonumber\\
 &&T(r)= \frac{8\epsilon\, M}{r^2}+\frac{-15120M^4-\epsilon[M^3(15120c_3-30240Mc_4)+\alpha M^2(80640+22680ln(2))+\beta(105840ln(2)+32768)]}{7560M^2 r^4}\nonumber\\
 &&
+O\Big(\frac{1}{r^5}\Big), \nonumber\\
&& R^{\mu \nu \lambda \rho}R_{\mu \nu \lambda \rho}=-\frac{48M^2}{r^6}+8\epsilon\Bigg[\frac{M^2 \alpha(19845ln(2)+70560)+\beta(92610ln(2)+28672)+M^3
(13230c_3-26460Mc_4)}{945Mr^7}\Bigg]+O\Big(\frac{1}{r^8}\Big),
\nonumber\\
&&
R^{\mu \nu}R_{\mu \nu}= O\Big(\epsilon^2 \Big)\,,
 \qquad R= \epsilon\Bigg[\frac{M^2 \alpha(2835ln(2)+10080)+\beta(13230ln(2)+4096)+M^3(1890c_3-3780Mc_4)}{945M^2r^4}\Bigg]+O\Big(\frac{1}{r^{5}}\Big).\nonumber\\
&&
  \end{eqnarray}
 Due to the non-linearity of the field equations (\ref{q8}), Eq. (\ref{invc}) do not coincides to Eq. (\ref{inv}) when the parameter $\beta=0$. This is similar to the black hole presented in \cite{Nashed:2018cth} in which the authors presented a  black hole solution for $f(T)=T+\frac{\alpha }{2} T^2+\frac{\beta}{3}T^3$ and when $\beta=0$ this black hole does not reduce to the black hole solution presented in \cite{Awad:2017tyz} for $f(T)=T+\frac{\alpha }{2}T^2$.

  Same discussion carried out for the quadratic solution given in Subsection \ref{S6336aa} can be applied to the invariants of the cubic case given by Eq. (\ref{invc}).

Using the same procedure of the quadratic form done for the calculation of the energy we calculated the energy of solution(\ref{pot}) from  Eqs. (\ref{sol2})
and (\ref{sol3}) and obtained the following:
 \begin{eqnarray} \label{Equd}
 E=M+\epsilon\frac{11340M^4c_4-3780M^3c_3-8505M^2ln(2)+\beta(16384-39690ln(2))}{7560 M^3}.
  \end{eqnarray}
  Equation (\ref{Equd}) reveals that $7560 M^4+\epsilon[11340M^4c_4-3780M^3c_3-8505M^2ln(2)+\beta(16384-39690ln(2))]>0$.

\begin{figure}
\centering
\includegraphics[scale=0.4]{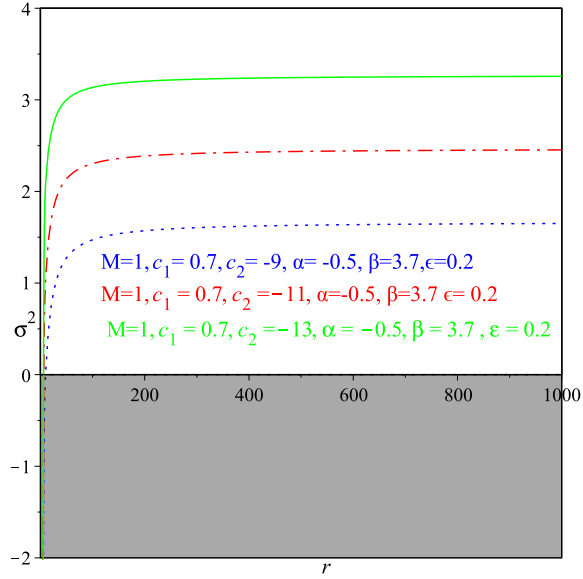}\hspace{0.2cm}
\caption{ {Schematic plot of  solution (\ref{pot}) using Eqs. (\ref{sol2}) and (\ref{sol3}), namely, $\sigma^2$  versus the coordinate, $r$.}}
\label{Fig:2}
\end{figure}

 Further, we calculated the geodesic deviation of  solution  (\ref{pot}) from Eqs. (\ref{sol2}) and (\ref{sol3}) and derived the condition of stability.
  This condition was quite  lengthy  and was not presented here. However,  Fig. \ref{Fig:2} shows its behavior for particular values of the model.
 This figure shows the regions where the BH solution are stable and those where there is no possible stability.
\section{Thermodynamics of ``p=2'' and cubic black holes }\label{S5}
In this section, we  investigated the thermodynamics behavior of the BH solutions (\ref{bfin}) and (\ref{pot}), from Eqs. (\ref{sol2}) and (\ref{sol3}),
 that are  related to the quadratic and cubic forms of the field equations (\ref{q8}).
To do this, we   gave the basic definitions of the thermodynamical quantities.
\subsection{Thermodynamics of ``p=2''}\label{S5a}
\begin{figure}
\centering
\subfigure[~Possible one horizon]{\label{fig:3a}\includegraphics[scale=0.3]{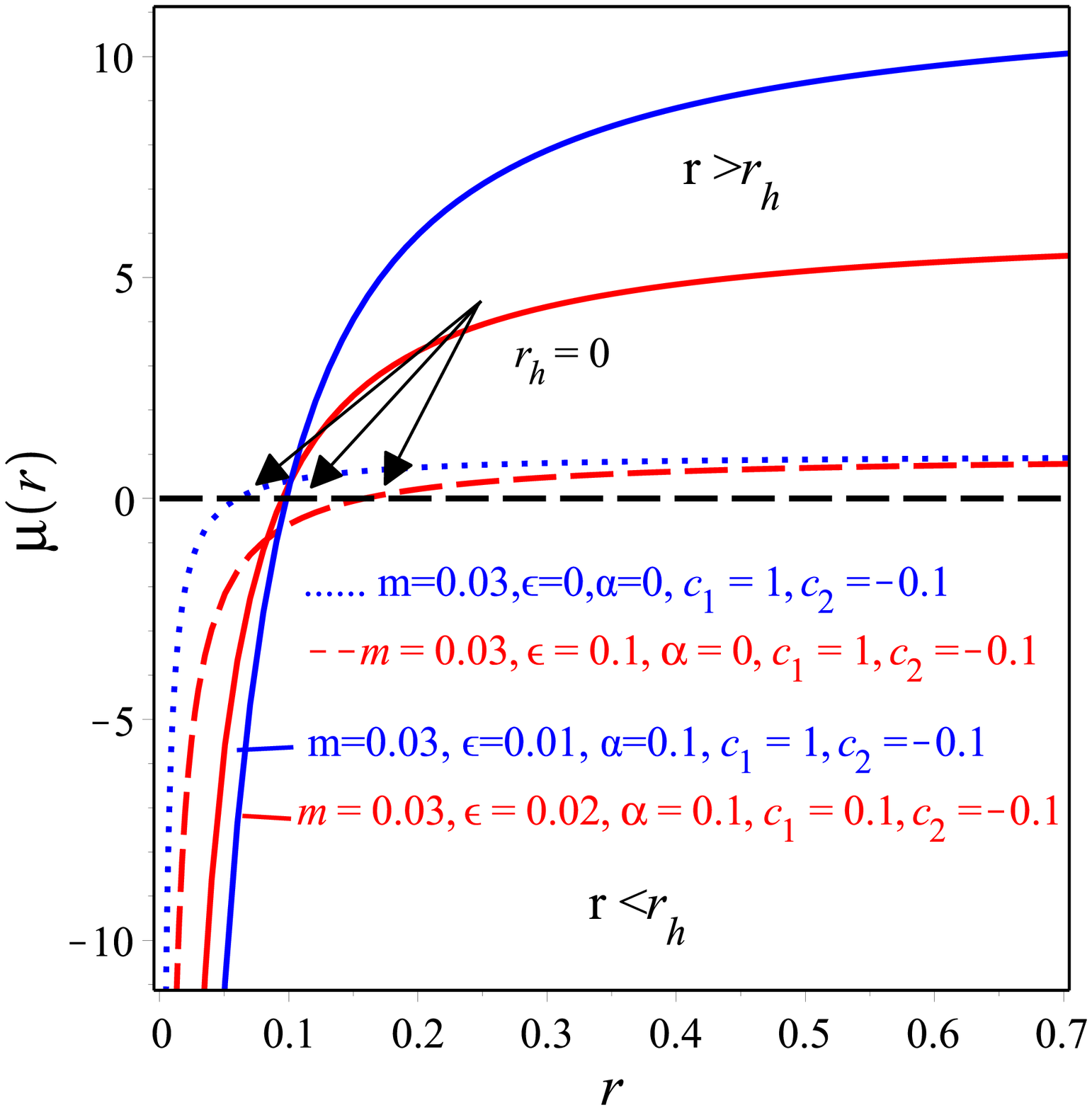}}
\subfigure[~The horizon mass-radius]{\label{fig:3b}\includegraphics[scale=0.3]{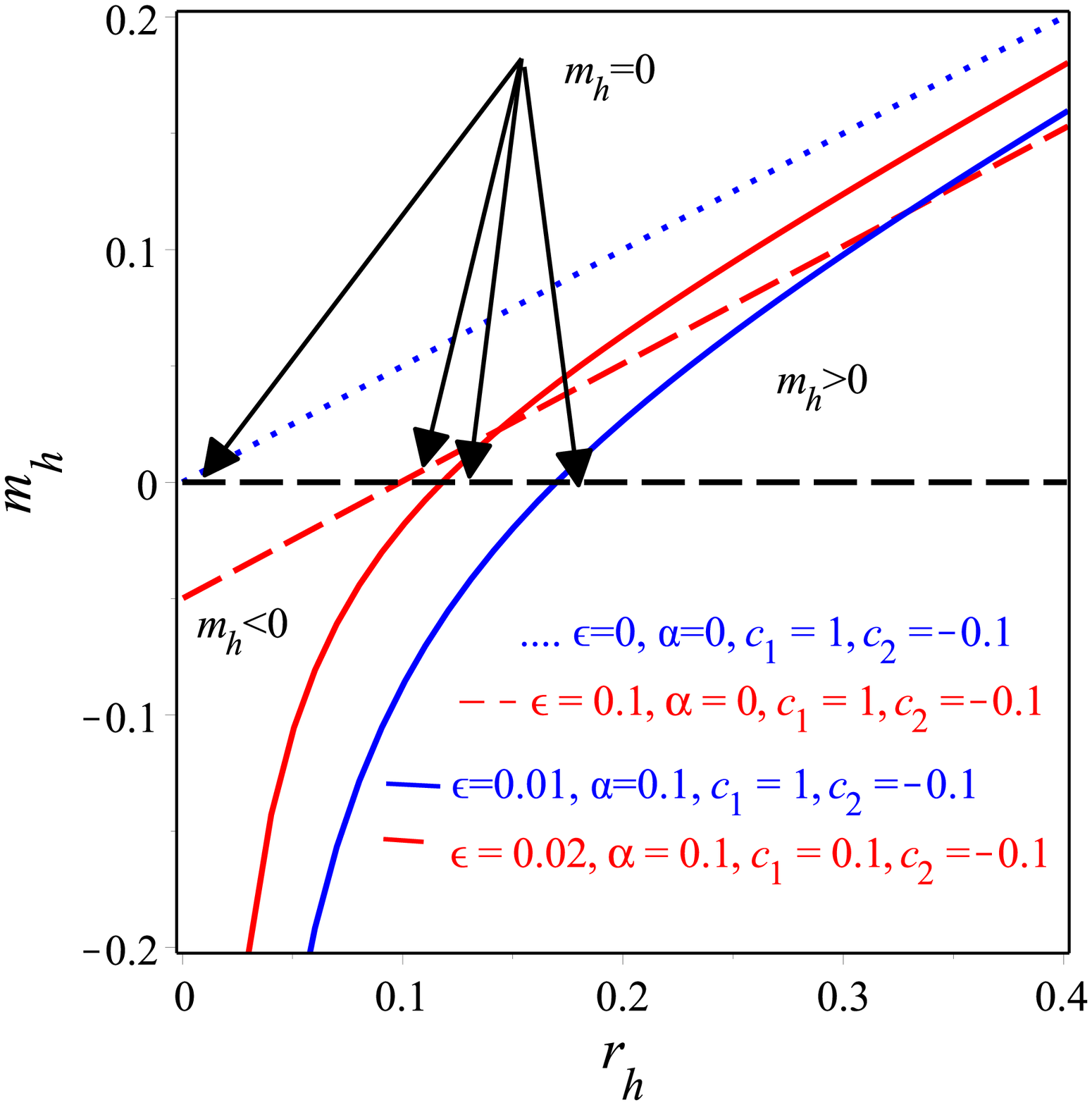}}
\subfigure[~The horizon Bekenstein-Hawking entropy]{\label{fig:3c}\includegraphics[scale=0.3]{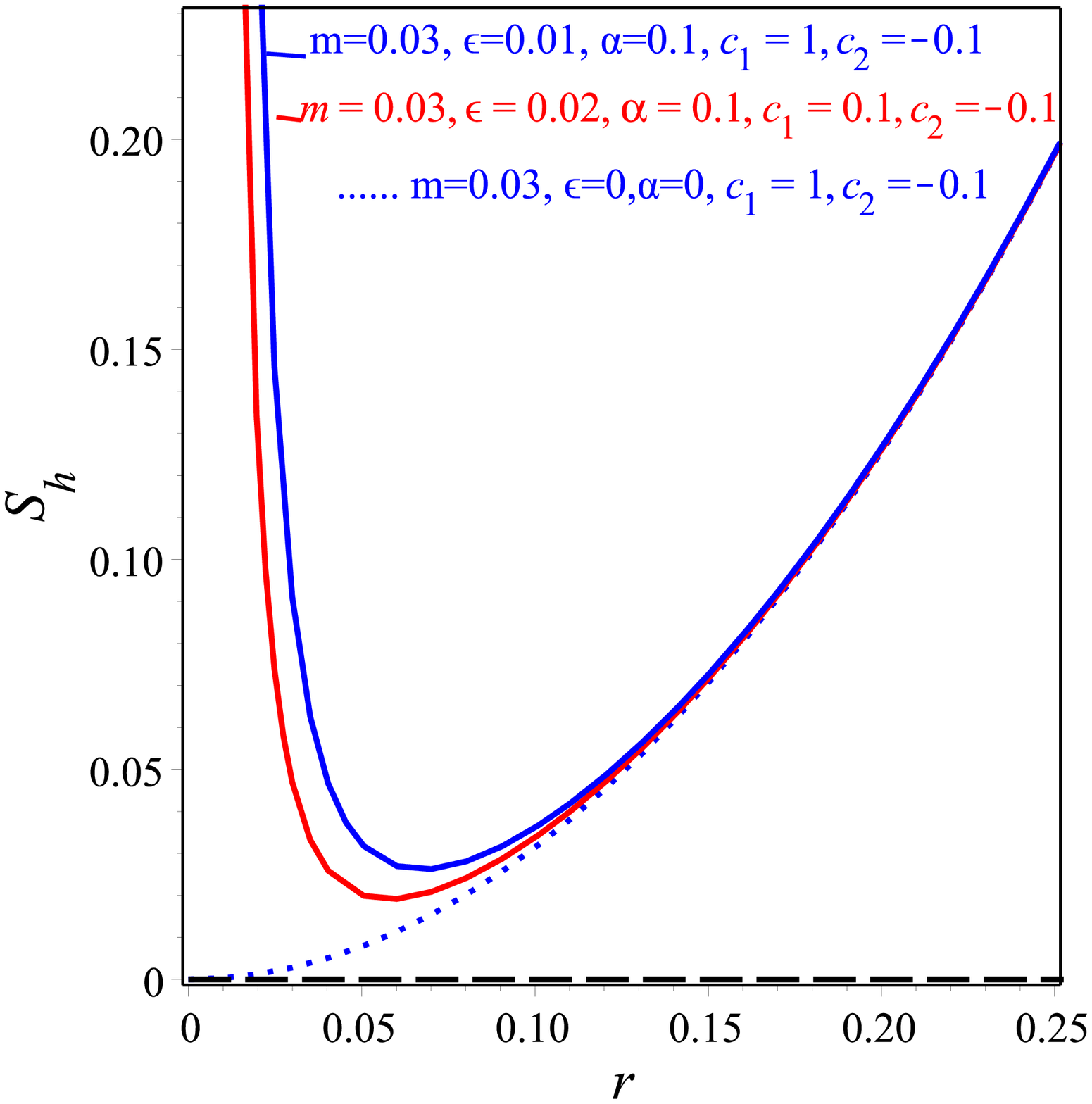}}\\
\subfigure[~The horizon Hawking Temperature]{\label{fig:3d}\includegraphics[scale=0.3]{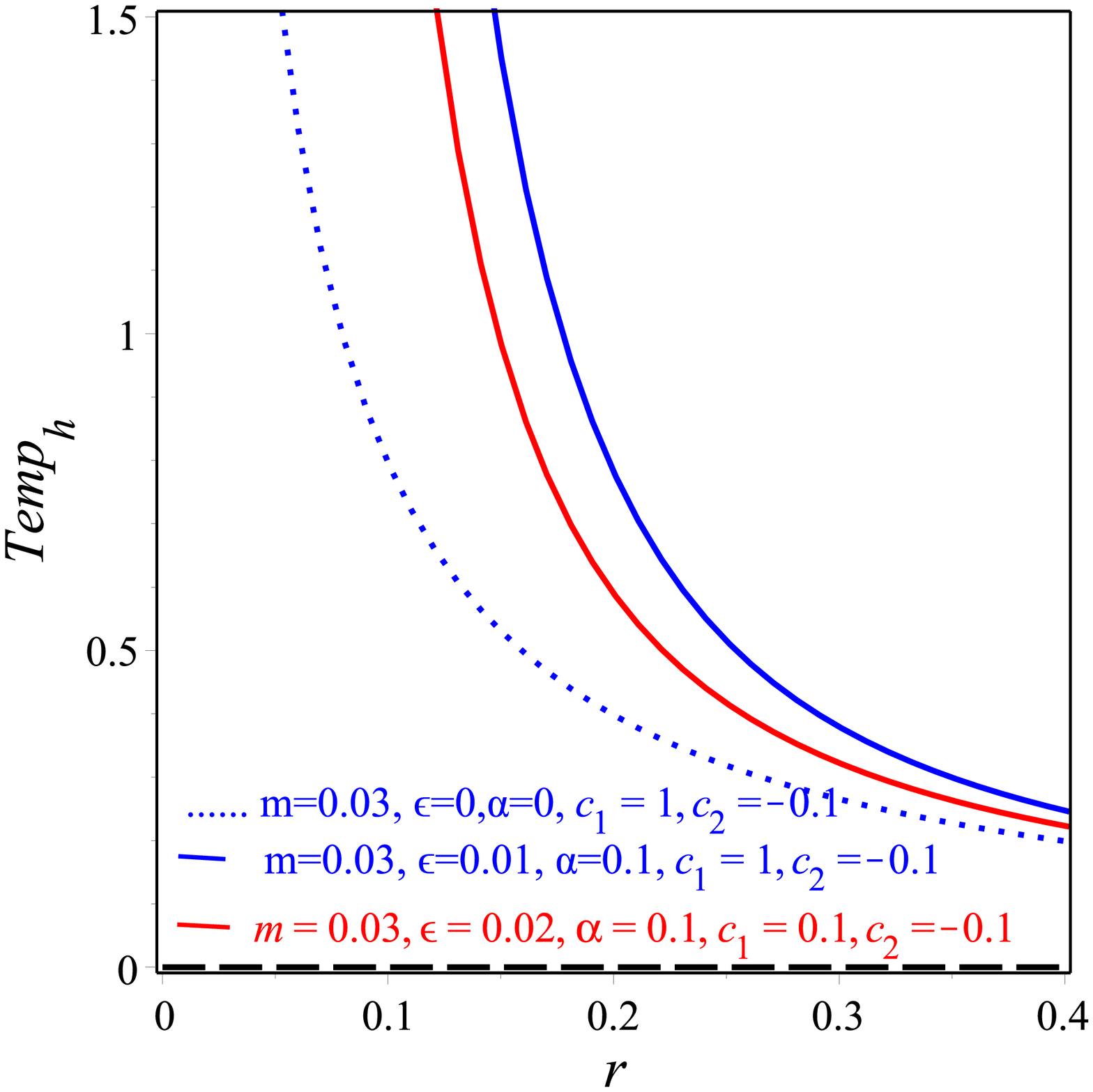}}
\subfigure[~The horizon heat capacity]{\label{fig:3e}\includegraphics[scale=0.3]{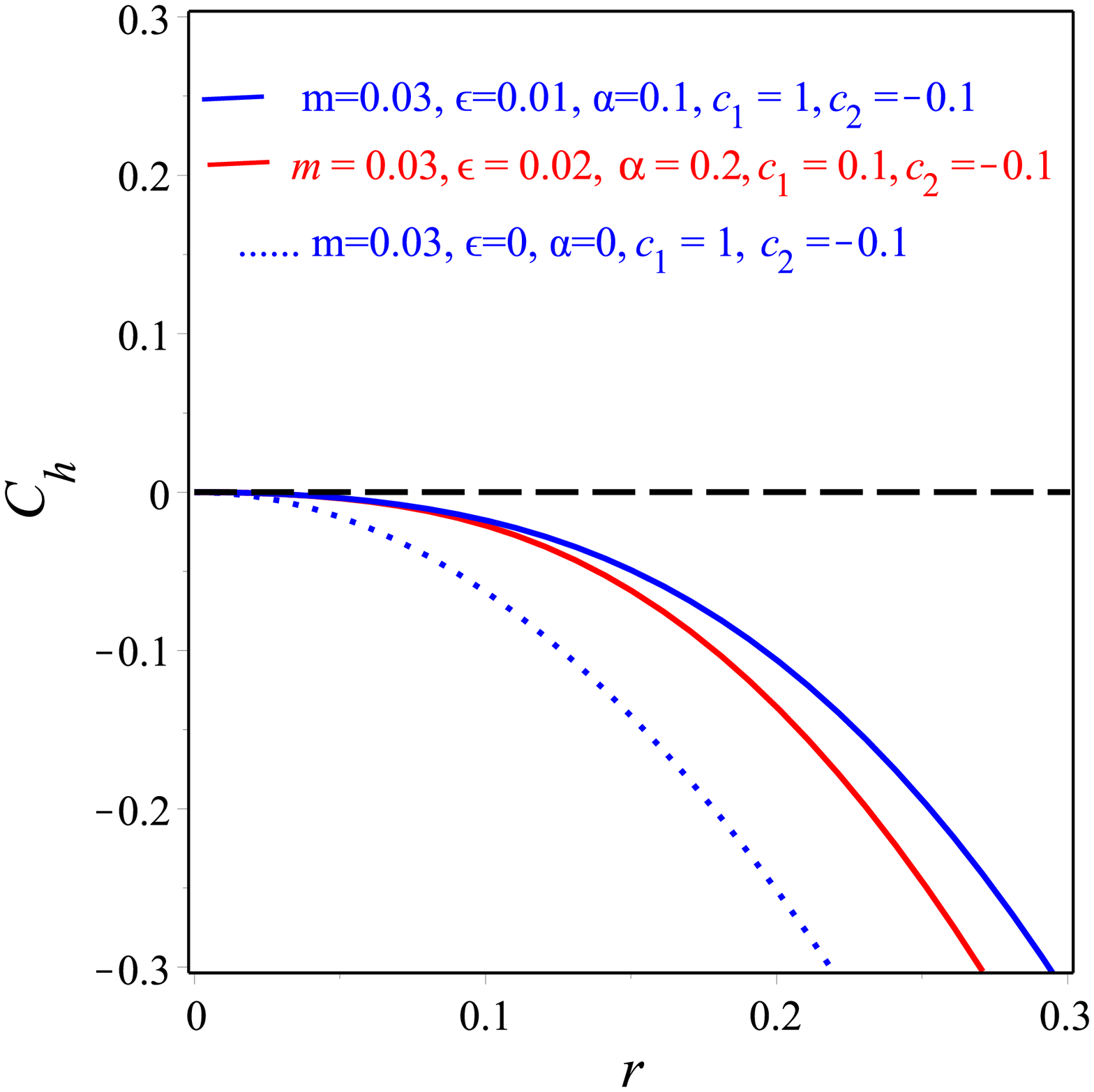}}
\subfigure[~The horizon Gibbs free energy]{\label{fig:3f}\includegraphics[scale=0.3]{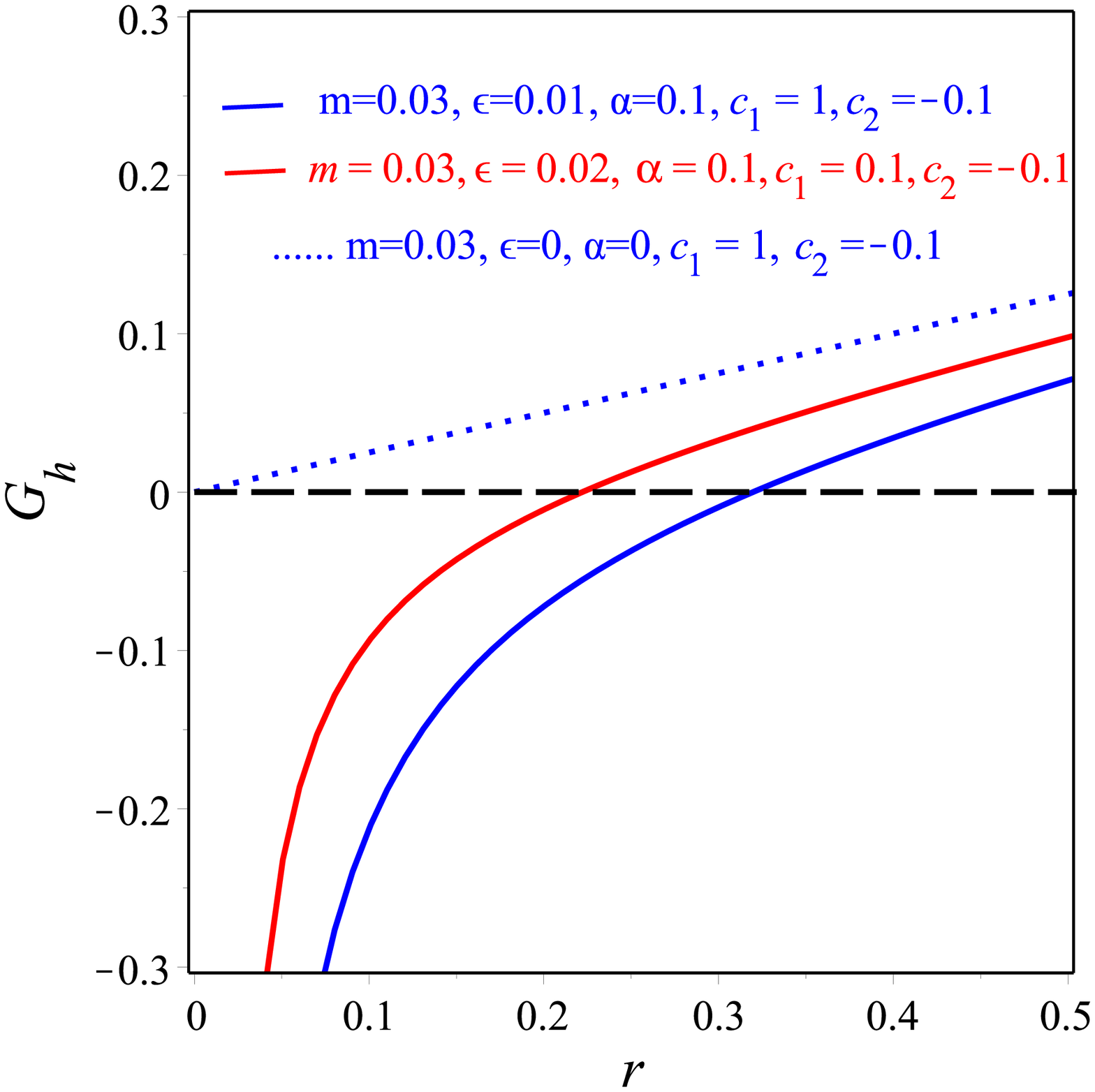}}
\caption{Schematic plots of the thermodynamical quantities of the BH solution (\ref{bfin}) for positive value of $\epsilon$: \subref{fig:3a}
 Typical behavior of the metric function $\mu(r)$ given by (\ref{bfin}); \subref{fig:3b} the horizon mass--radius relation (\ref{hor-mass-rad1a});
  \subref{fig:3c}
   Typical behavior of the horizon entropy, which shows that $S_h$ increases quadratically as $r_h$ increases and we have always a positive entropy;
   \subref{fig:3d}
    Typical behavior of the horizon temperature, (\ref{m44-ee}),  and \subref{fig:3e} the heat capacity, (\ref{heat-cap1a}),
     which show that both vanish at $r_{h}$, and we have always positive temperature and negative heat capacity which shows that
the BH is unstable;
     \subref{fig:3f} Typical behavior of the horizon Gibbs free energy which shows that $G_h$   have
    negative value as $r<r_h$ and  positive value as $r>r_h$ for positive $\epsilon$.}
\label{Fig:3}
\end{figure}

The metric potential of the temporal component of Eq. (\ref{bfin}) takes the following form:
 \begin{eqnarray} \label{hor}
&& \mu(r)= 1- \frac{2m}{r}+\epsilon\left[-\frac{c_1}{r}+c_2-\alpha\left(\frac{m^2+6 m r+r^2}{m r^3}-\frac{16 \left(1-\frac{2 m}{r}\right)^{3/2}}
    {3 m^2}+\frac{(1-\frac{3 m}{r})}{2 m^2} \ln \left(1-\frac{2m}{r}\right)\right)\right]\nonumber\\
    &&\simeq 1+\epsilon\Bigg[ c_2+\frac{\alpha\{64-9ln(2)\}}{12m^2}\Bigg]
    -\frac{1}{r}\Bigg[2m+\epsilon\Bigg(c_1+\frac {16\alpha}{m}\Bigg)\Bigg]-O\Big(\frac{1}{r^5}\Big)\,.\nonumber\\
\end{eqnarray}
Equation (\ref{hor}) was draw  in Fig. \ref{Fig:3}\subref{fig:3a}, the plot indicates  the one horizon of the BH which can be obtained in  a precise form from  the solution  of Eq. $\mu(r)=0$. This horizon is known as the event  horizon $r_h$.
One can calculate the total mass contained  $r_h$. This can be done by setting $\mu(r_h) = 0$,
and then we obtain the horizon mass-radius relation in the following form
\begin{eqnarray} \label{hor-mass-rad1a}
 {m_h}=\frac{3r_h{}^2+\epsilon(3r_h{}^2c_2-32\alpha-3r_hc_1)-9\epsilon \alpha ln(2)}{6r_h}\, .
\end{eqnarray}
Equation (\ref{hor-mass-rad1a}) is plotted  in Fig. \ref{Fig:3}\subref{fig:3b}, whereas $m_h$ has  positive and negative values, the BH has one horizon when
$m_h=m_{min}$. This result is consistent with Fig. \ref{Fig:3}\subref{fig:3a}.

The Hawking temperature is usually defined as \cite{PhysRevD.86.024013,Sheykhi:2010zz,Hendi:2010gq,PhysRevD.81.084040}
  \begin{equation}
T_h = \frac{\mu'(r_h)}{4\pi},
\end{equation}
where the event horizon $r = r_h$ is the positive solution of the equation  $\mu(r_h) = 0$ which satisfies $\mu'(r_h)\neq 0$.
In the framework of $f(T)$ gravity, the entropy   is given by \cite{PhysRevD.84.023515,Zheng:2018fyn}
\begin{equation}\label{ent}
S(r_h)=\frac{1}{4}Af_{T}(r_h),
\end{equation}
where $A$  represents the  area.

The constraint $\mu(r_h) = 0$ yields
\begin{eqnarray} \label{m33}
&&   {r_h}_{{}_{{}_{{}_{{}_{``p=2"}}}}}\simeq\frac{12m^2+\epsilon[(6mc_1+32\alpha-12c_2m^2)+9\alpha ln(2)]}{6m}\,.
\end{eqnarray}
When $\epsilon=0$ we get the GR limit.

From Eq. (\ref{ent}), the entropy of solution (\ref{bfin})  takes the form
\begin{eqnarray} \label{ent1}
{S_h}_{{}_{{}_{{}_{{}_{``p=2"}}}}}=\frac{\pi[r_h{}^5+2\epsilon\alpha m^2(4m+3r_h)]}{r_h{}^3}\,,
\end{eqnarray}
which shows that when $\epsilon=0$ we get the GR entropy. Equation (\ref{ent1}) shows that the parameters $\epsilon$ and $\alpha$  should be either positive
 or negative to get positive entropy otherwise the entropy will have a negative quantity. The behaviour of Eq.  (\ref{ent1}) is shown in
  Fig. \ref{Fig:3}\subref{fig:3c} for positive values of   $\epsilon$ and $\alpha$  which shows positive value of entropy.

The Hawking temperatures of solution (\ref{bfin}) takes the form,
\begin{equation}\label{m44-ee}
{T_h}_{{}_{{}_{{}_{{}_{``p=2"}}}}}\simeq\frac{3r_h{}^2+\epsilon[\alpha(64-9ln(2))+3c_2r_h{}^2]}{12\pi r^3}\,.
\end{equation}
Equation (\ref{m44-ee}) shows that when $\epsilon=0$ we get the Hawking temperature of Schwarzschild BH. We  depicted the Hawking's
temperature in  Fig. \ref{Fig:3}\subref{fig:3d} for positive values of   $\epsilon$ and $\alpha$.   Fig. \ref{Fig:3}
\subref{fig:3d}  proves that we do
have a   positive temperature for the BH  (\ref{bfin}) and the temperature   may take a  negative value when either $\epsilon$ or $\alpha$
become a negative.

The stability of the BH solution is an important topic that can be studied on the dynamical and the perturbative levels
 \cite{Nashed:2003ee,Myung_2011,Myung:2013oca}.  To investigate the thermodynamical stability of BH solution one derives the formula of the
 heat capacity $C(r_h)$
  at the event horizon. The event horizon heat capacity is given by the following from \cite{Nouicer:2007pu,DK11,Chamblin:1999tk}:
\begin{equation}\label{heat-capacity}
C_h\equiv C(r_h)= \frac{\partial m_h}{\partial T}=\frac{\partial m_h}{\partial r_h} \left(\frac{\partial T}{\partial r_h}\right)^{-1}\, .
\end{equation}
The BH will be thermodynamically stable, if its heat capacity $C_h$ is positive, and will be unstable if $C_h$ is negative.
Using (\ref{hor-mass-rad1a}) and (\ref{m44-ee}) into (\ref{heat-capacity}), we obtain the heat capacity as
\begin{equation}\label{heat-cap1a}
{C_h}_{{}_{{}_{{}_{{}_{``p=2``}}}}}=\frac{2\pi r_h^2 }{3}\frac{3r_h{}^2+\epsilon[3r_h{}^2c_2+9\alpha ln(2)+32\alpha]}
{\epsilon[9\alpha ln(2)-r_h{}^2c_2-64\alpha]-r_h{}^2} \, .
\end{equation}
Equation  (\ref{heat-cap1a}) shows that $C_h$ does not locally diverge and the BH has no phase transition of second-order.
The heat capacity is depicted in Fig. \ref{Fig:3}\subref{fig:3e} which shows that $C_h<0 $ where $r_h<r_{dg}$ and the BH
 is thermodynamically unstable. The main reason that makes the heat capacity negative is the derivative of Hawking
  temperature and this is consistent with the nature of Schwarzschild black hole which can be discovered when $\epsilon=0$.
  In the non-vanishing of $\epsilon$ we can create a positive heat capacity but the price of this
is to accept the Hawking temperature to has a negative value.

\begin{figure}
\centering
\subfigure[~Possible one horizon]{\label{fig:4a}\includegraphics[scale=0.3]{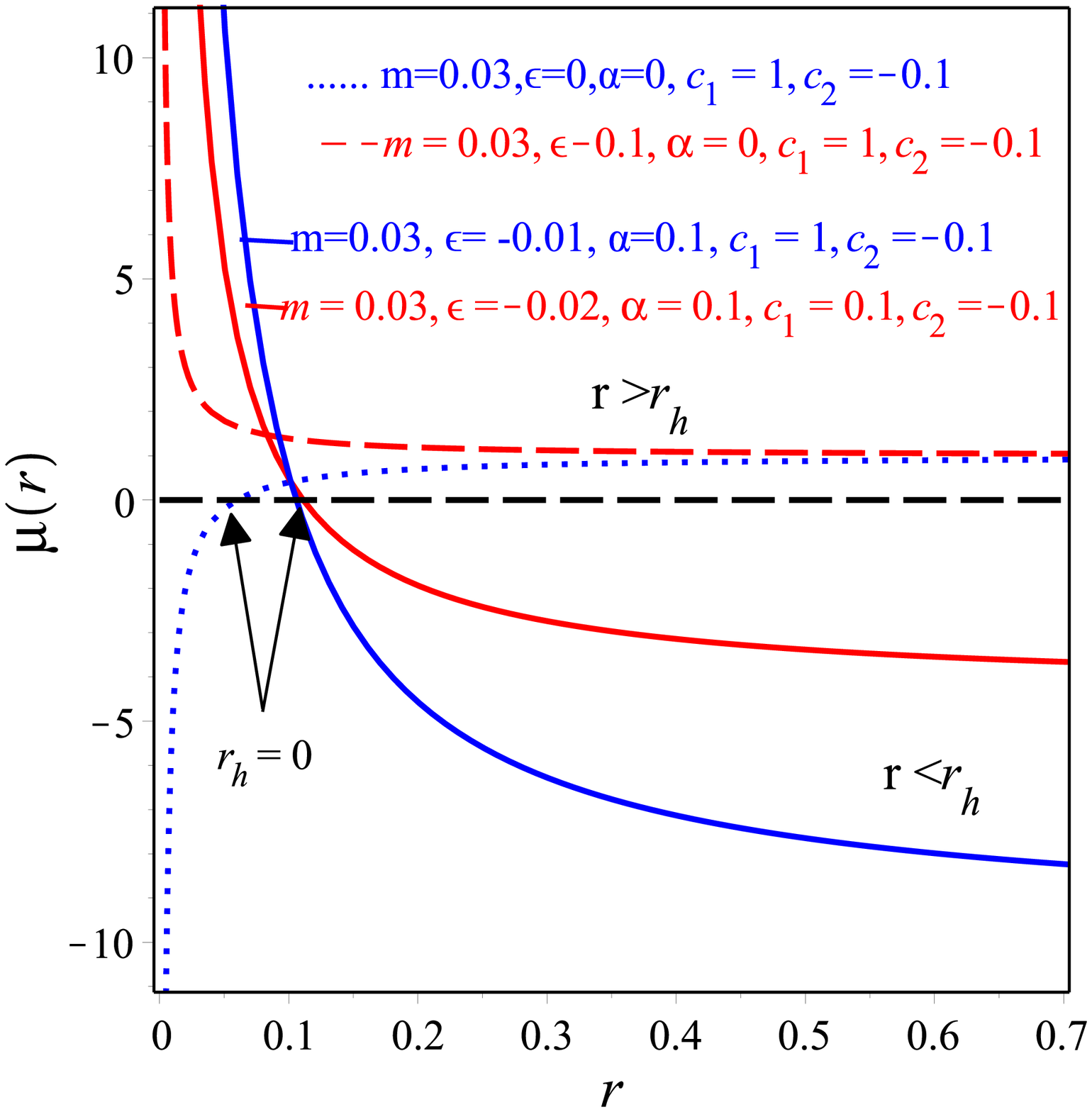}}
\subfigure[~The horizon mass-radius]{\label{fig:4b}\includegraphics[scale=0.3]{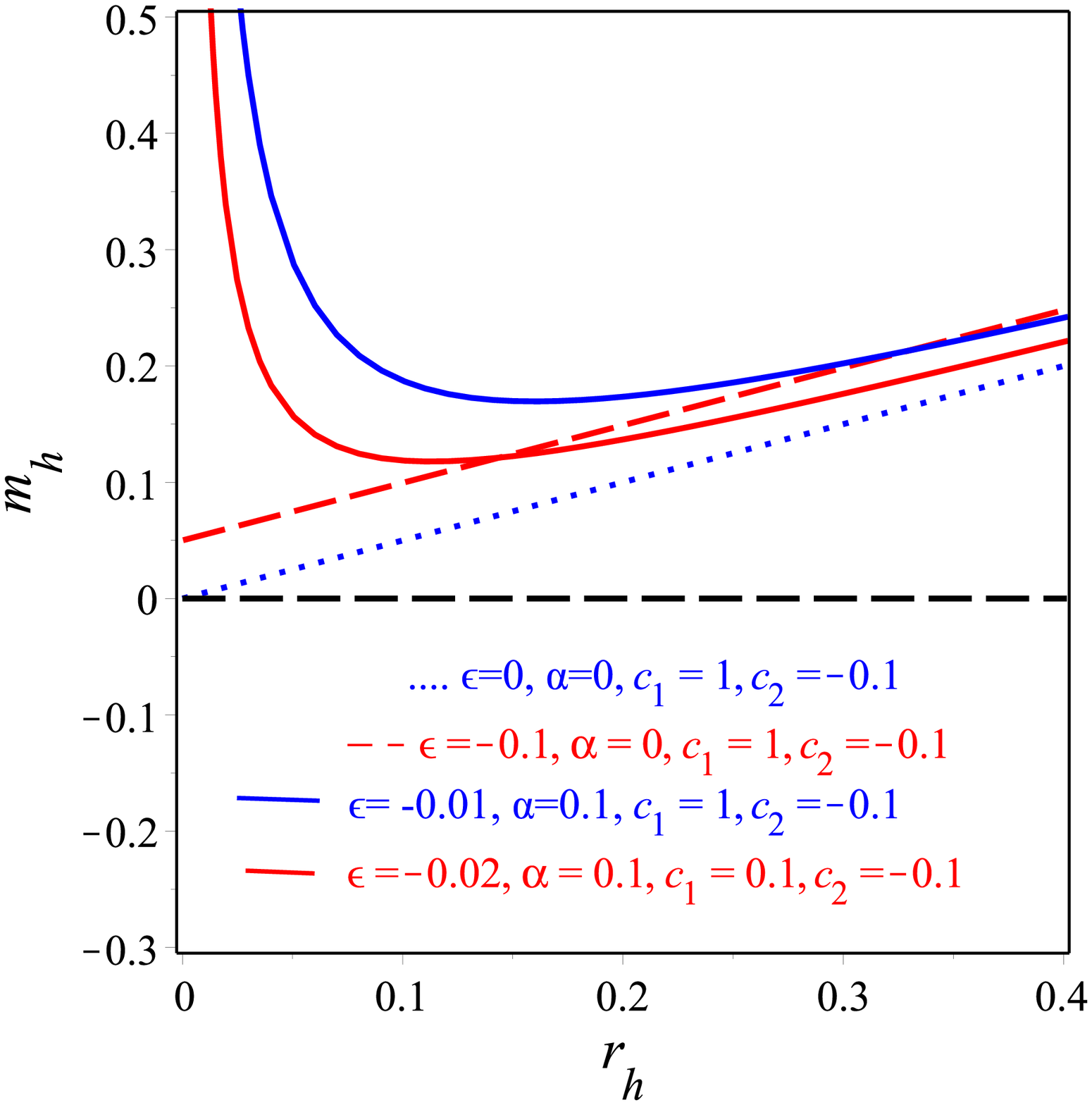}}
\subfigure[~The horizon Bekenstein-Hawking entropy]{\label{fig:4c}\includegraphics[scale=0.3]{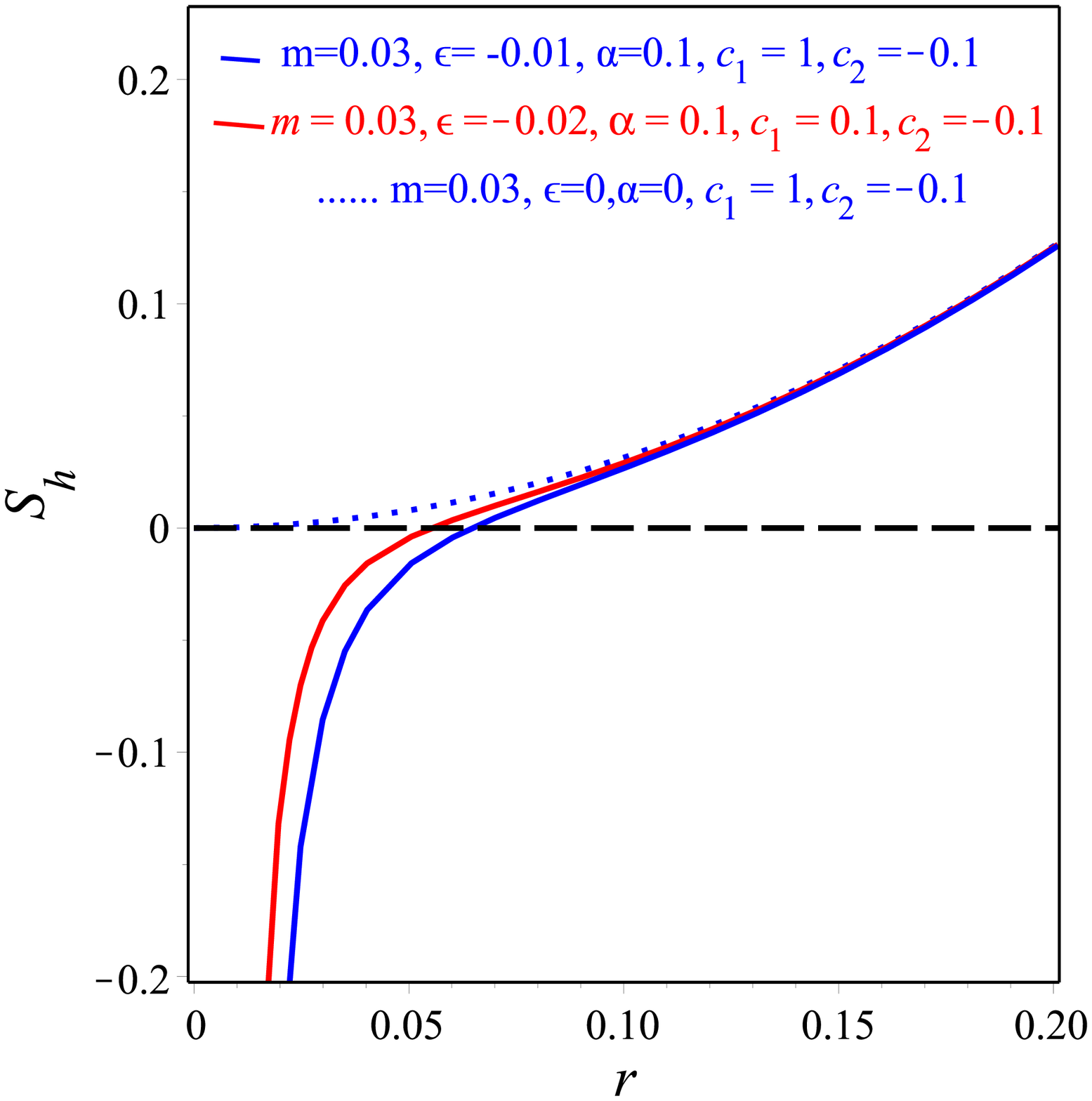}}\\
\subfigure[~The horizon Hawking Temperature]{\label{fig:4d}\includegraphics[scale=0.3]{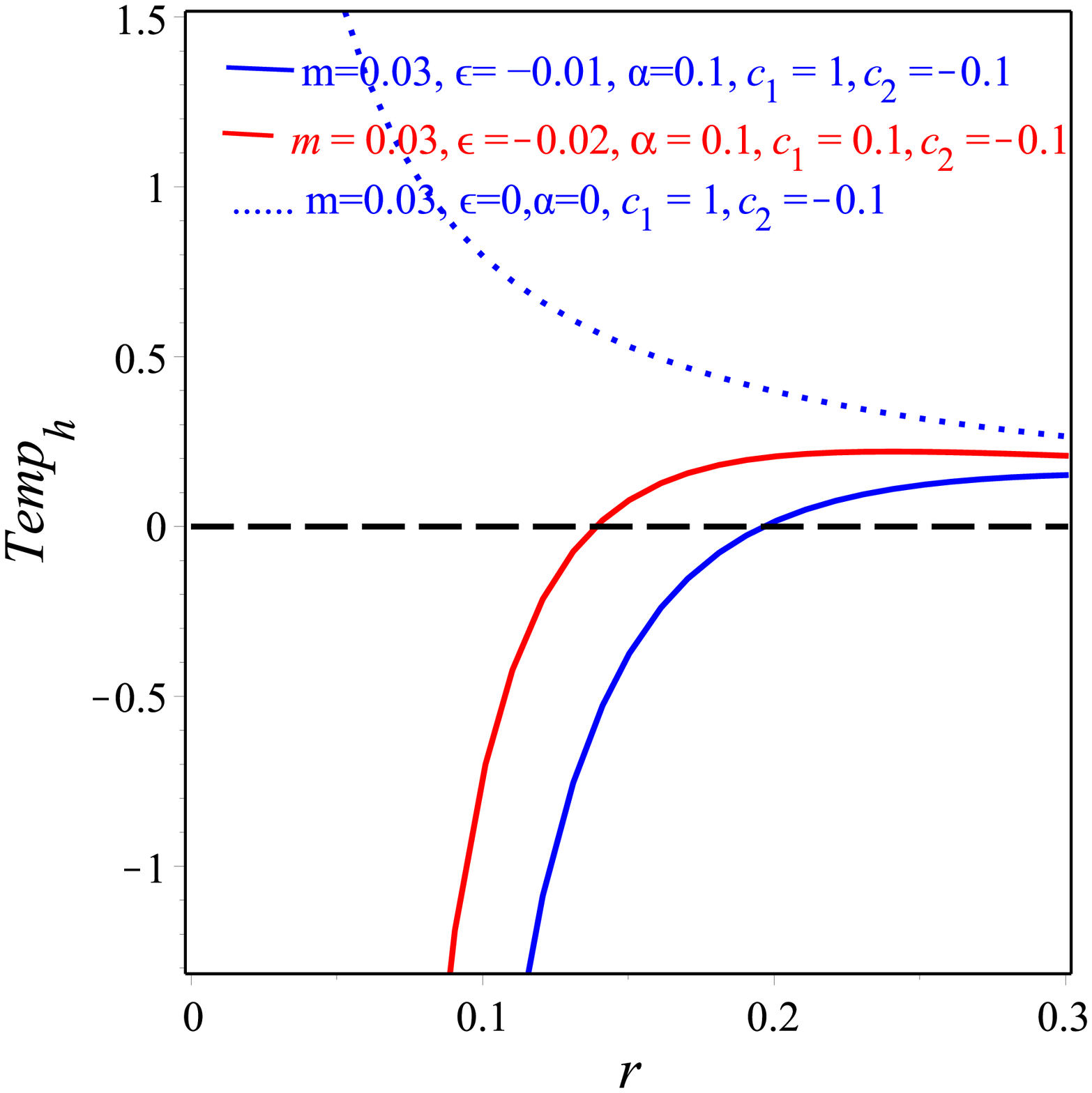}}
\subfigure[~The horizon heat capacity]{\label{fig:4e}\includegraphics[scale=0.3]{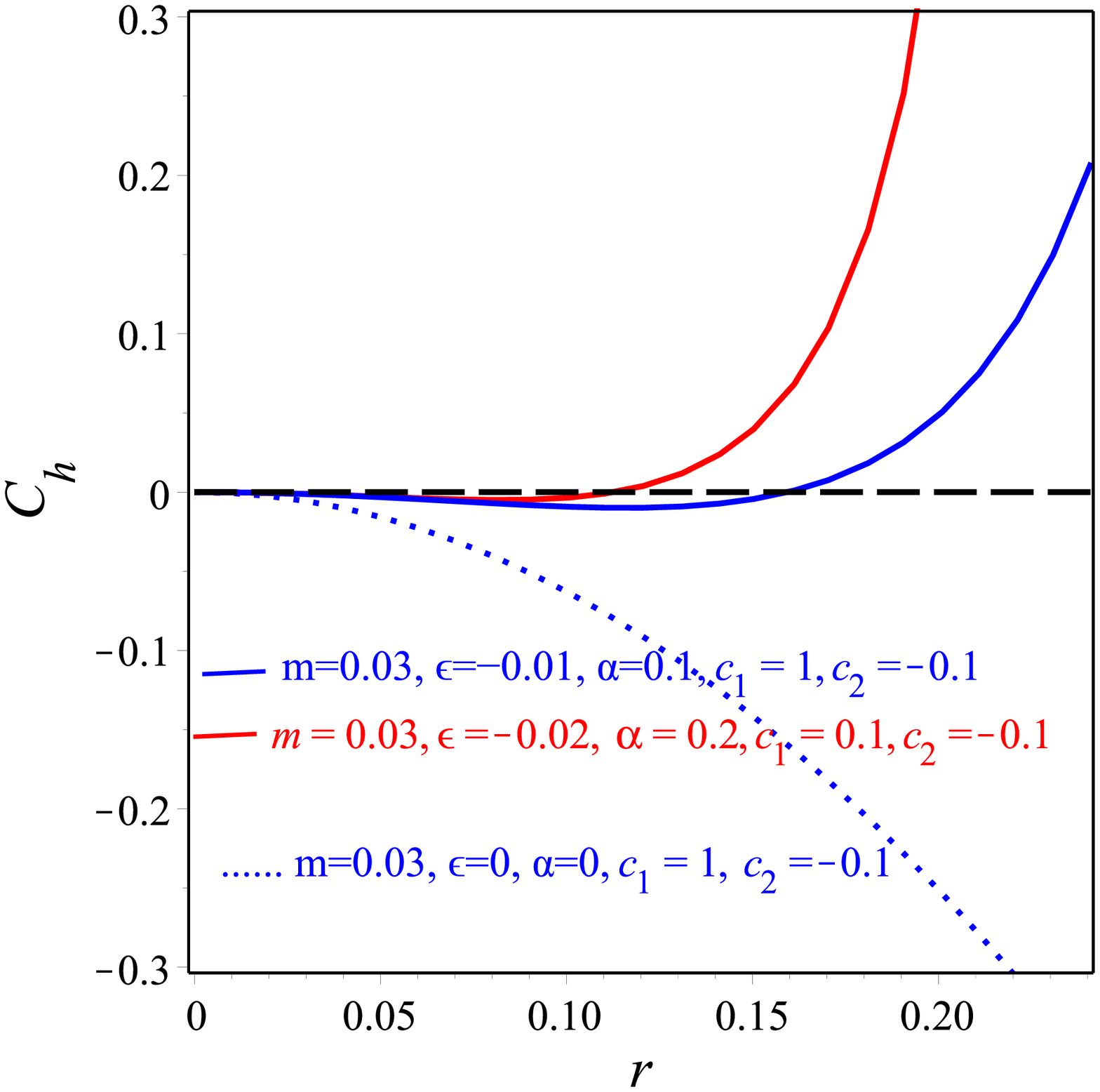}}
\subfigure[~The horizon Gibbs free energy]{\label{fig:4f}\includegraphics[scale=0.3]{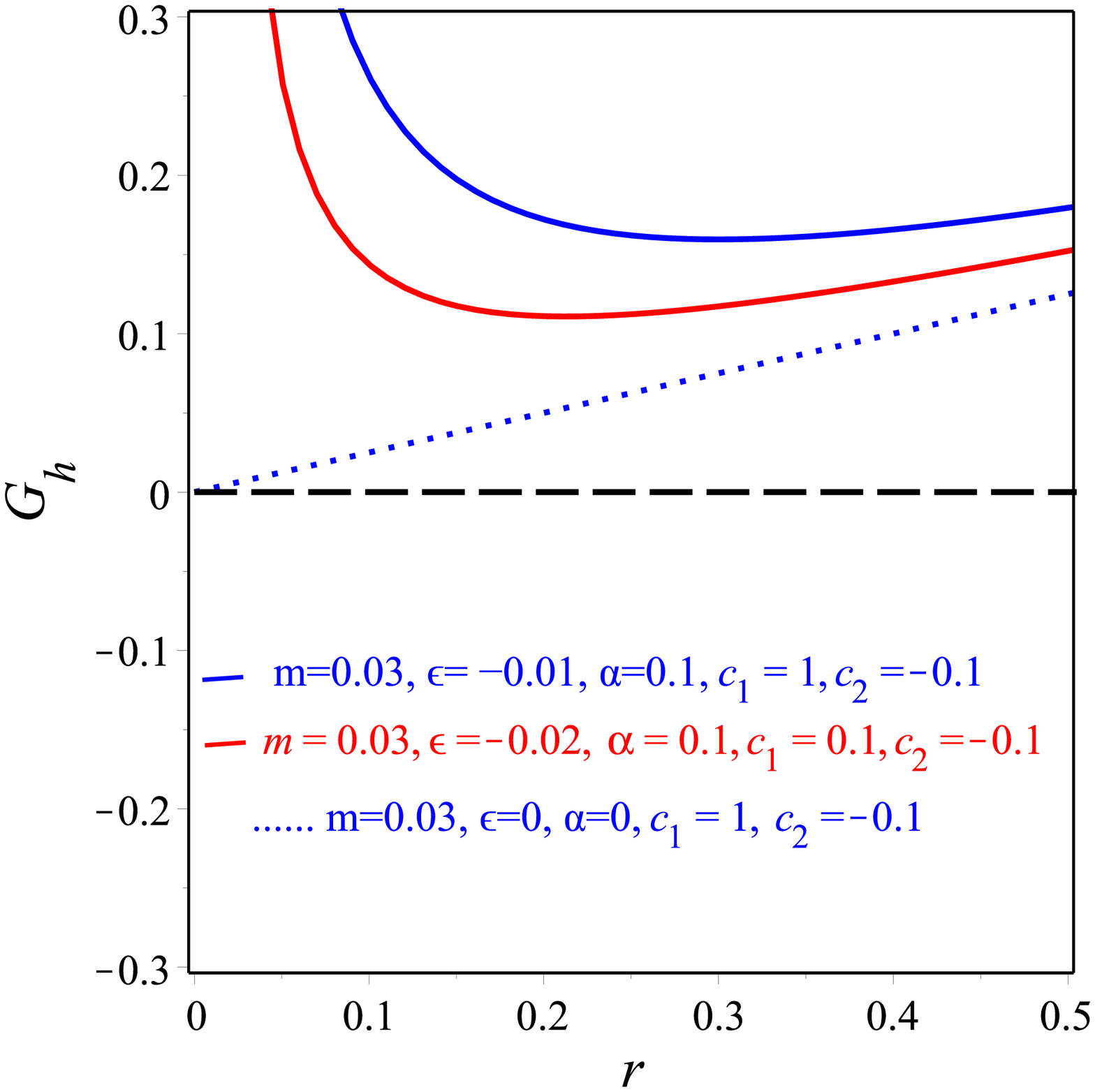}}
\caption{Schematic plots of thermodynamical quantities of the BH solution (\ref{bfin}) for negative value of $\epsilon$: \subref{fig:4a}
 Typical behavior of the metric function $\mu(r)$ given by (\ref{bfin}); \subref{fig:4b} the horizon mass--radius relation (\ref{hor-mass-rad1a}); \subref{fig:4c}
   typical behavior of the horizon entropy, which shows that $S_h$ increases quadratically as $r_h$ increases and as $r<r_h$ we have a
   negative entropy and as $r>r_h$ we have a positive entropy; \subref{fig:4d}
    typical behavior of the horizon temperature,  and \subref{fig:4e} the heat capacity, (\ref{heat-cap1a}),
     which shows that both vanish at $r_{h}$, and
    as $r<r_{h}$ we have a
   negative temperature and heat capacity and as $r>r_{h}$ we have a positive values of both
the BH is unstable (i.e $C_h<0$) when $r<r_{h}$ and it is stable (i.e $C_h>0$) when $r>r_{h}$;
     \subref{fig:4f} typical behavior of the horizon Gibbs free energy which shows that $G_h$ could have
     only positive values for negative value of $\epsilon$.}
\label{Fig:4}
\end{figure}

The  Gibbs free energy is given by \cite{Zheng:2018fyn,Kim:2012cma}
\begin{equation} \label{enr1}
G(r_h)=m(r_h)-T(r_h)S(r_h)\,.
\end{equation} The quantities
 $m(r_h)$, $T(r_h)$ and $S(r_h)$  are the mass, temperature and entropy  at the event horizon, respectively.  From Eqs. (\ref{hor-mass-rad1a}),
(\ref{ent1}) and (\ref{m44-ee}) in (\ref{enr1}), we obtain
\begin{eqnarray} \label{m77}
&&{G_h}_{{}_{{}_{{}_{{}_{\tiny Eq. (\ref{bfin})}}}}}=\frac{3r_h{}^5+\epsilon[3r_h{}^5c_2-9\alpha r_h{}^3ln(2)-128\alpha r_h{}^3-6r_h{}^4 c_1-6m^2\{3\alpha r_h+4
\alpha m^2\}]}{12r_h{}^4}.
\end{eqnarray}
We depict the Gibbs energy of the black hole (\ref{bfin}) in Fig.  \ref{Fig:3}\subref{fig:3f},   which indicates that the Gibbs energy has negative values
 at large $r<r_h$ and positive value when $r>r_h$. We note that for $\epsilon=0$, the Schwarzschild black hole is recovered which is shown in Fig.
  \ref{Fig:3}\subref{fig:3f} by the blue dot curve. Interestingly, for the negative value of $\epsilon$, Gibbs energy is always positive, and
  as we discuss before that the price of this is the negative Hawking temperature. We depict the case of a negative
value of $\epsilon$ in Fig. \ref{Fig:4}.

\subsection{Thermodynamics of the cubic solution}\label{S5b}
\begin{figure}
\centering
\subfigure[~Possible one horizon]{\label{fig:5a}\includegraphics[scale=0.3]{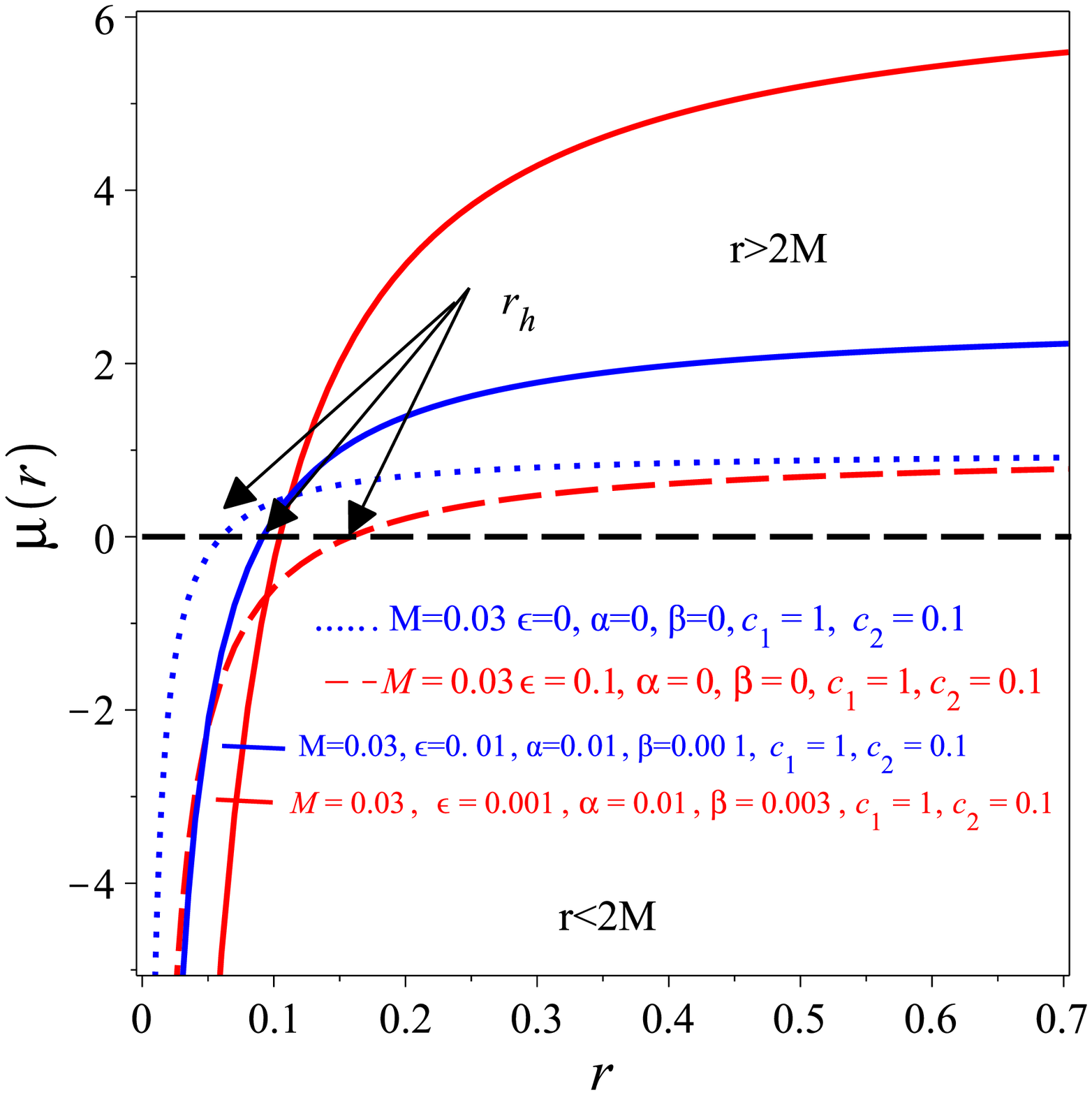}}
\subfigure[~The horizon mass-radius]{\label{fig:5b}\includegraphics[scale=0.3]{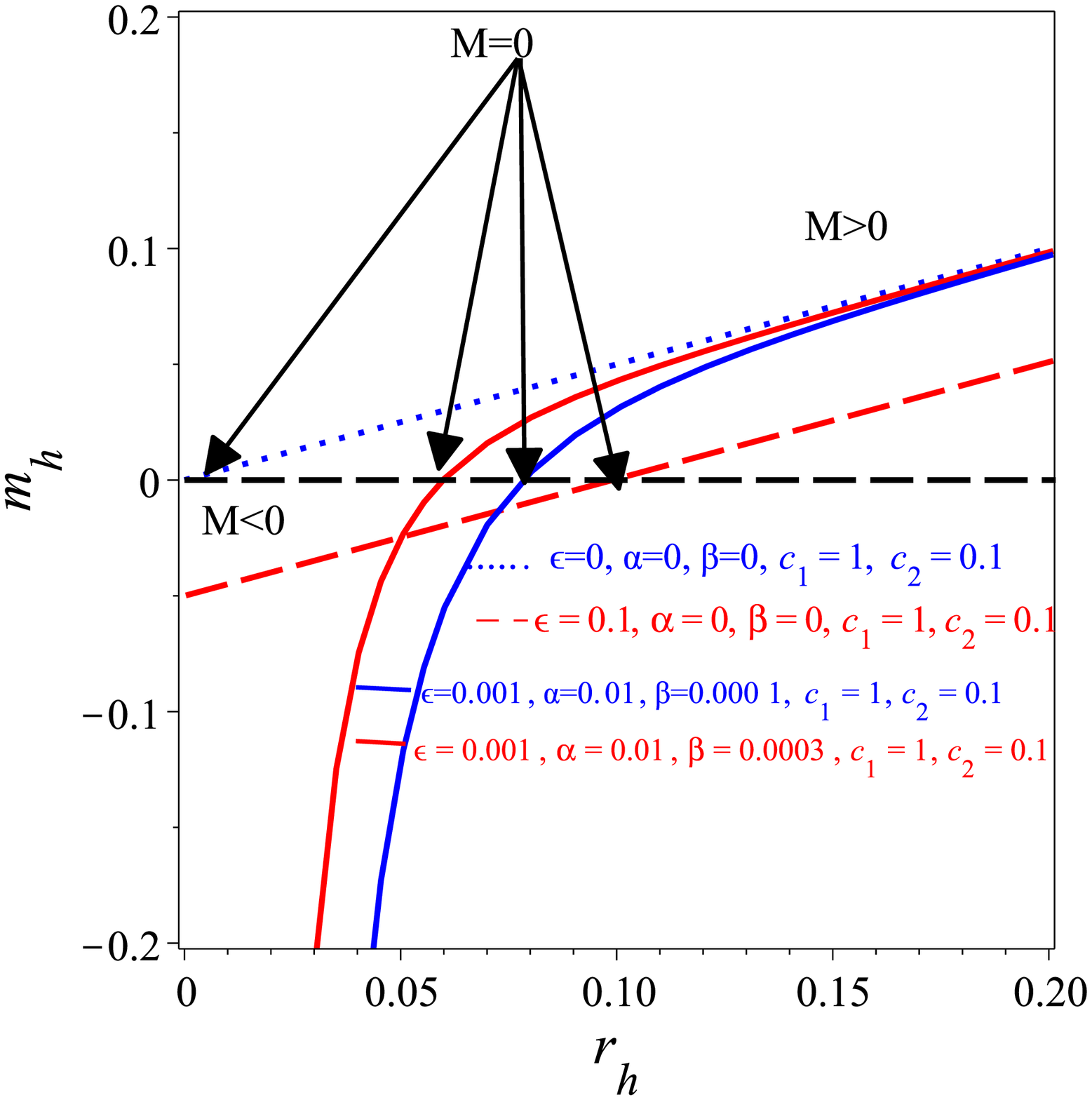}}
\subfigure[~The horizon Bekenstein-Hawking entropy]{\label{fig:5c}\includegraphics[scale=0.3]{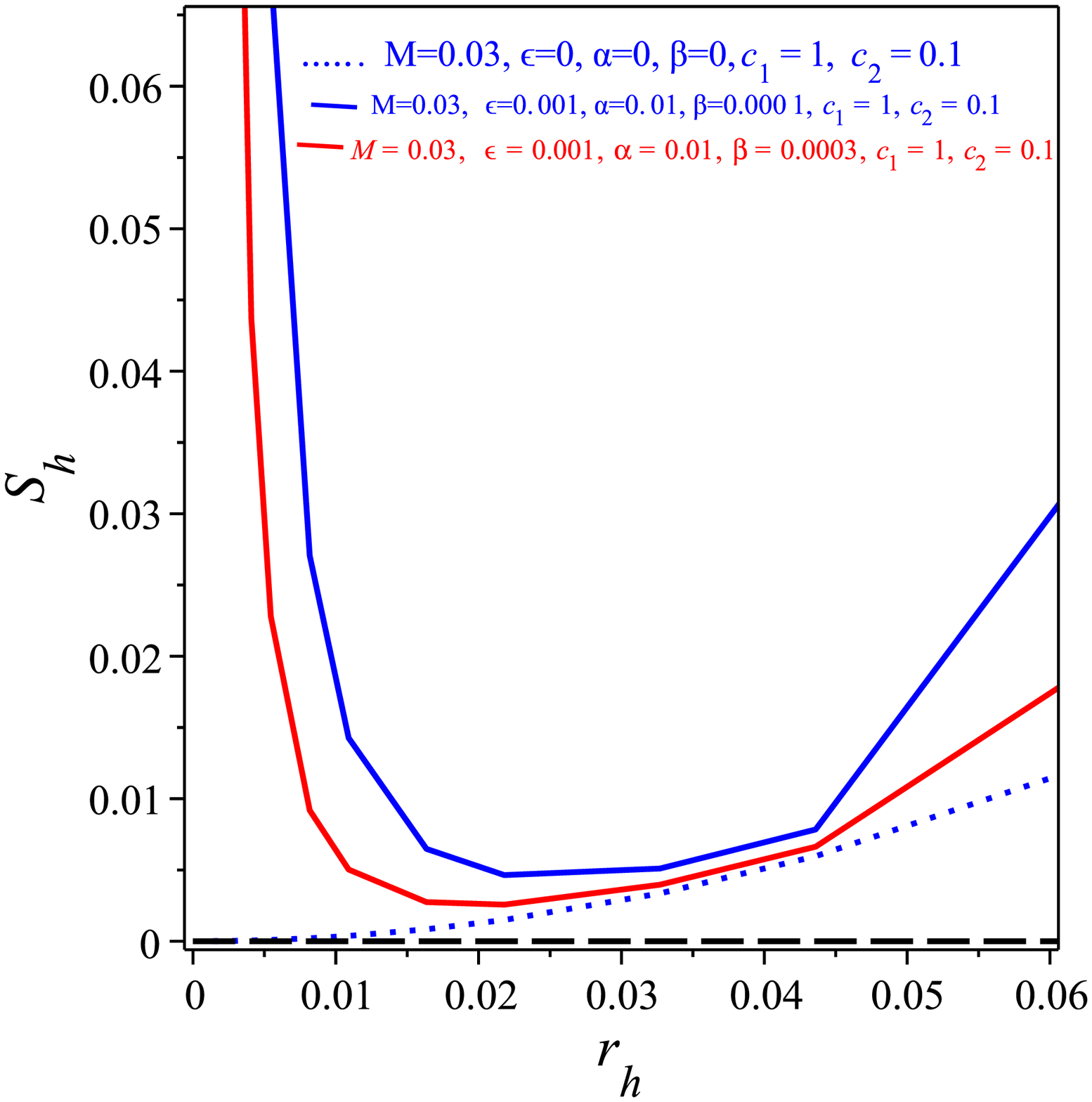}}\\
\subfigure[~The horizon Hawking Temperature]{\label{fig:5d}\includegraphics[scale=0.3]{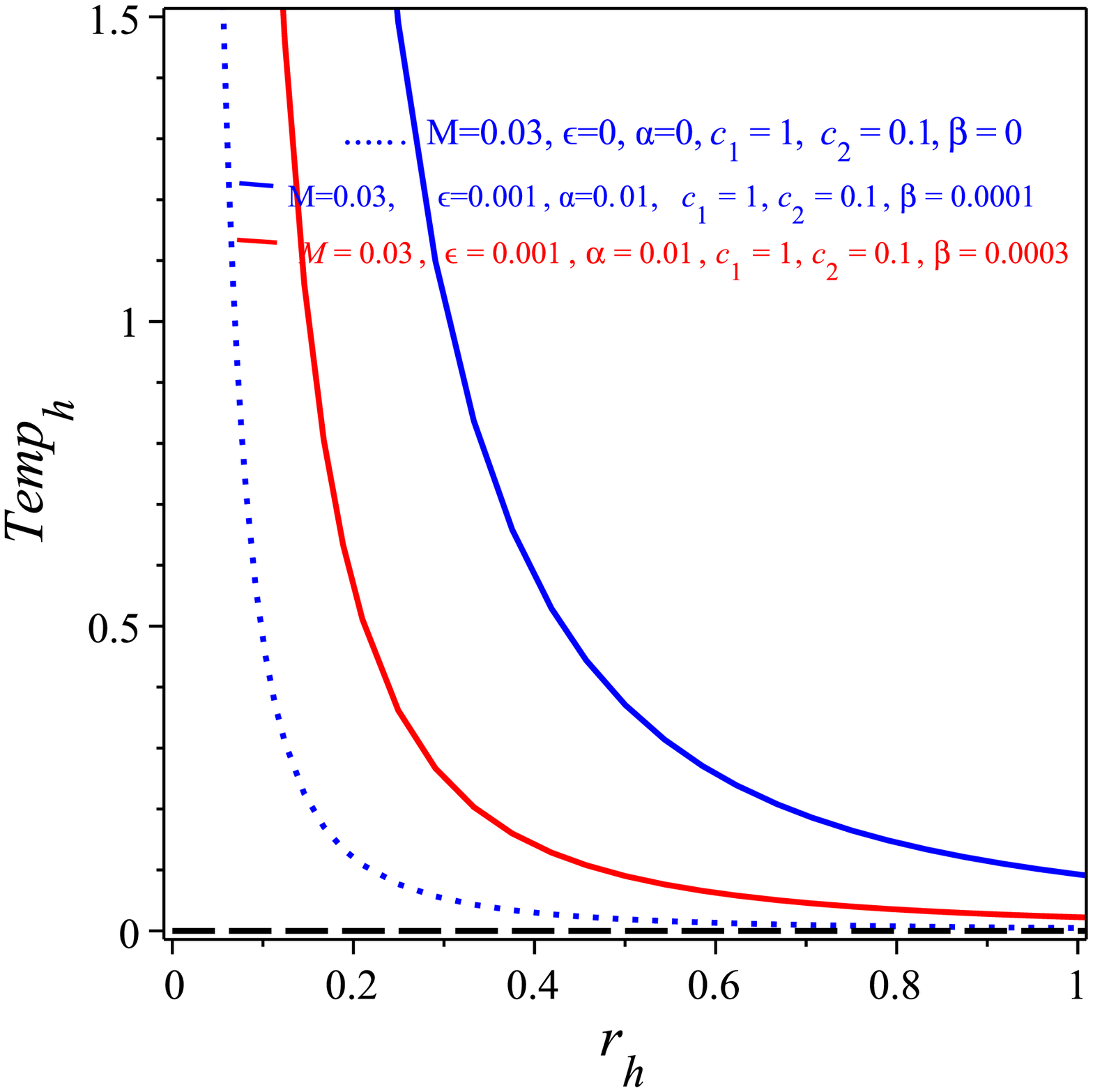}}
\subfigure[~The horizon heat capacity]{\label{fig:5e}\includegraphics[scale=0.3]{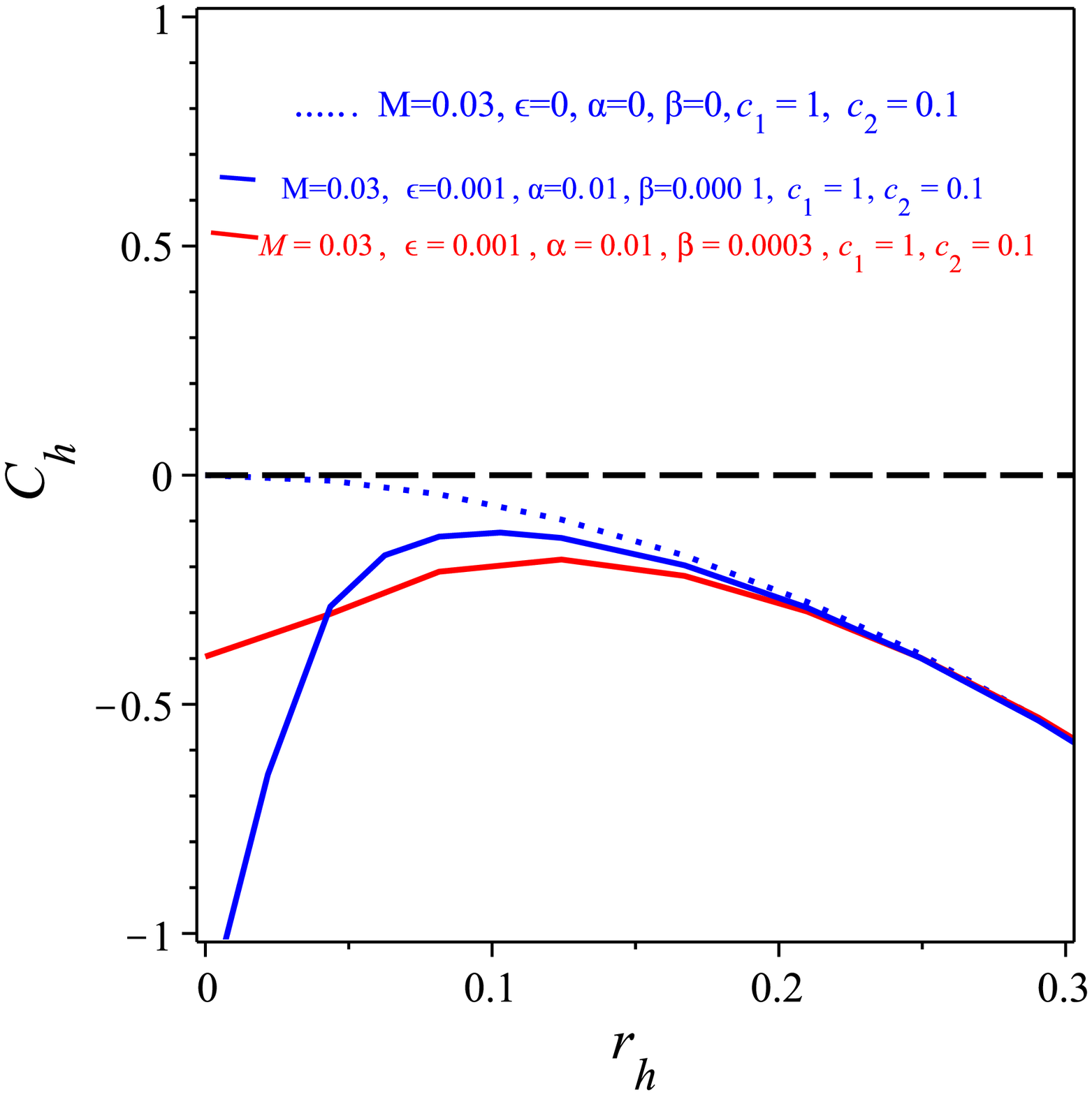}}
\subfigure[~The horizon Gibbs free energy]{\label{fig:5f}\includegraphics[scale=0.3]{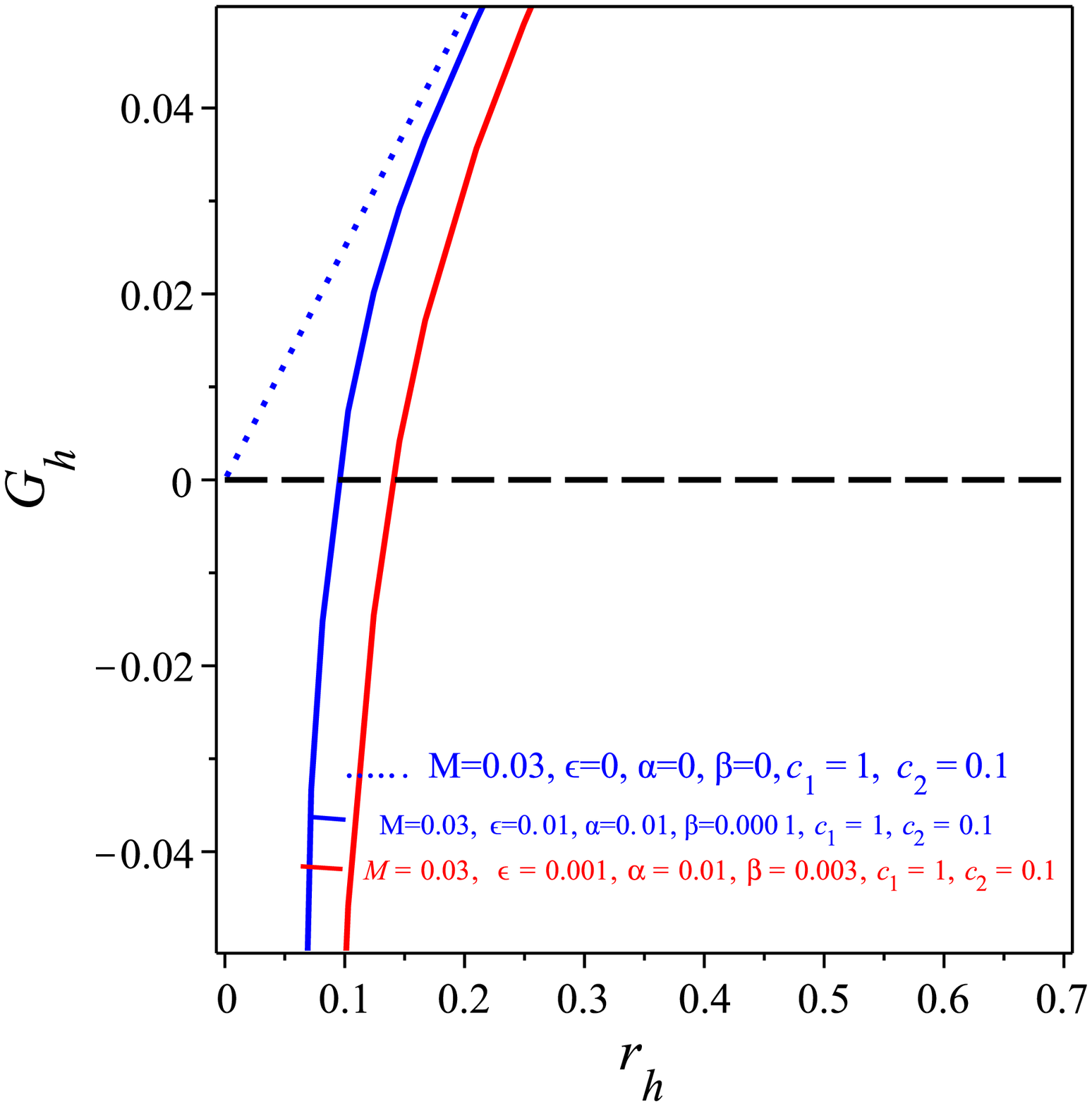}}
\caption{Schematic plots of thermodynamical quantities of the BH solution (\ref{pot}), using Eqs. (\ref{sol2}) and (\ref{sol3})
 for positive value of $\epsilon$: \subref{fig:5a}
 Typical behavior of the metric function $\mu(r)$ given by  (\ref{horc11}); \subref{fig:5b}
  the horizon mass--radius relation (\ref{hor-mass-rad1a});
  \subref{fig:5c}
   typical behavior of the horizon entropy, which shows that $S_h$ increases quadratically as $r_h$ increases and we have always a positive entropy;
   \subref{fig:5d}
    typical behavior of the horizon temperature,  and \subref{fig:5e} the heat capacity, (\ref{heat-cap1a}),
     which show that both vanish at $r_{h}$, and
    we have always positive temperature and negative heat capacity;
     \subref{fig:5f} typical behavior of the horizon Gibbs free energy which shows that $G_h$ could have
         negative value at $r<r_h$ and  positive value at $r>r_h$  for negative value of $\epsilon$.}
\label{Fig:5}
\end{figure}

The metric potential of the temporal component of solution (\ref{pot}), using Eqs. (\ref{sol2}) and (\ref{sol3}) takes the form
 \begin{eqnarray} \label{horc11}
&& \mu(r)\simeq 1+\epsilon\Bigg[ \frac{1260m^4c_2-\alpha m^2\{945ln(2)ln(2) -6720\}-\beta\{4410ln(2)+8192\}}{1260m^4}\Bigg]\nonumber\\
&& -\frac{1}{135m^3r}\Bigg[270m^4+\epsilon\Bigg(135m^3c_1+2160\alpha m^2+2048\beta\Bigg)\Bigg]-O\Big(\frac{1}{r^5}\Big)\,.\nonumber\\
\end{eqnarray}
Equation (\ref{horc11}) is drawn  in Fig. \ref{Fig:5}\subref{fig:3a}, the plot shows that the BH could have one horizon
 at the root of $\mu(r)=0$, this is the  event  horizon $r_h$.
The horizon mass-radius relation of solution (\ref{pot}), using Eqs. (\ref{sol2})  and (\ref{sol3}) has the form
\begin{eqnarray} \label{hor-mass-rad1ac}
 {m_h}=\frac{135r_h{}^4+\epsilon[135r_h{}^4c_2-405\alpha r_h{}^2-135r_h{}^3c_1-405n \alpha ln(2)  r_h{}^2-16348 \beta]}{720r_h{}^3}\, .
\end{eqnarray}
We plot the above relation in Fig. \ref{Fig:5}\subref{fig:5b}, whereas $m_h$ has  positive and negative values, the black hole has one horizon when
$m_h=m_{min}$. This result is in agreement with Fig. \ref{Fig:5}\subref{fig:5a}.

The constraint $\mu(r_h) = 0$ yields
\begin{eqnarray} \label{m33c}
&&   {r_h}_{{}_{{}_{{}_{{}_{cubic}}}}}\simeq\frac{3780m^4+\epsilon[4096\beta+10080\alpha m^2+1890c_1m^3+13230\beta ln(2)+2835\alpha m^2ln(2)-3780c_2m^4]}{1890m^3}\,.
\end{eqnarray}
When $\epsilon=0$ Eq. (\ref{m33c}) gives  the GR limit.

From Eq. (\ref{ent}), the entropy of solution (\ref{pot})  takes the form
\begin{eqnarray} \label{ent1c}
&&{S_h}_{{}_{{}_{{}_{{}_{cubic}}}}}=\frac{\pi}{r_h{}^2(r_h{}^2-4m^2)^2}\Bigg[r_h{}^4(r_h{}^2-4m^2)^2+4\epsilon\Big\{r_h{}^6\alpha+16\beta m^4
+8\alpha r_h{}^2m^4
-6\alpha r_h{}^4m^2+8\beta m^3 r_h\nonumber\\
&& +\Big\{8[\beta m-\alpha r_h]  r_h{}^2-m[16\beta m^2-\alpha r_h{}^4]
+4r_h m^2[4\beta - mr_h\alpha]+r_h{}^3\alpha [4m^2- r_h{}^2]\Big\}\sqrt{r_h{}^2+2mr_h}\nonumber\\
&&+4\beta r_h{}^2[2r_h{}^2-7m^2]\Bigg]\,,
\end{eqnarray}
which shows that when $\epsilon=0$ we recover the entropy of GR. Equation (\ref{ent1c}) shows that the parameters $\epsilon$, $\alpha$ and $\beta$
 should be either positive
 or negative to get positive entropy otherwise the entropy will have a negative quantity. The behavior of Eq.  (\ref{ent1c}) is shown in
  Fig. \ref{Fig:5}\subref{fig:5c} for positive values of   $\epsilon$ and $\alpha$  which shows positive value of entropy.

The Hawking temperatures of solution (\ref{pot}) takes the form,
\begin{eqnarray} \label{m44ec}
{T_h}_{{}_{{}_{{}_{{}_{cubic}}}}}\simeq\frac{3r_h{}^2+\epsilon[\alpha(64-9ln(2))+3c_2r_h{}^2]}{12\pi r_h{}^3}\,,
\end{eqnarray}
which is identical with the quadratic form up to $\epsilon$ but with different value of $\alpha$ to unify the values through all the calculations of cubic case.
Therefore, same discussions carried out for the case of quadratic can be applied here.

Using (\ref{hor-mass-rad1ac}) and (\ref{m44ec}) into (\ref{heat-capacity}), we obtain the heat capacity as
\begin{equation}\label{heat-cap1ac}
{C_h}_{{}_{{}_{{}_{{}_{cubic}}}}}=\frac{2\pi  }{45}\frac{45r_h{}^4+\epsilon[45r_h{}^4c_2+135r_h{}^2\alpha ln(2)+480 r_h{}^2\alpha+16348\beta]}
{\epsilon[9\alpha ln(2)-r_h{}^2c_2-64\alpha]-r_h{}^2} \, .
\end{equation}
Equation  (\ref{heat-cap1ac}) shows that $C_h$ does not locally diverge and the black hole has no phase transition of second-order.
The heat capacity is depicted in Fig. \ref{Fig:5}\subref{fig:5e} which shows that $C_h<0 $ where $r_h<r_{dg}$ and the black hole
 is thermodynamically unstable. The main reason that makes the heat capacity negative is the derivative of Hawking temperature and this is consistent with the
 Schwarzschild black hole which can be discovered when $\epsilon=0$. In the non-vanishing of $\epsilon$ we can create a positive heat capacity but the price
 of this is to accept the Hawking temperature to has a negative value.

\begin{figure}
\centering
\subfigure[~Possible one horizon]{\label{fig:6a}\includegraphics[scale=0.3]{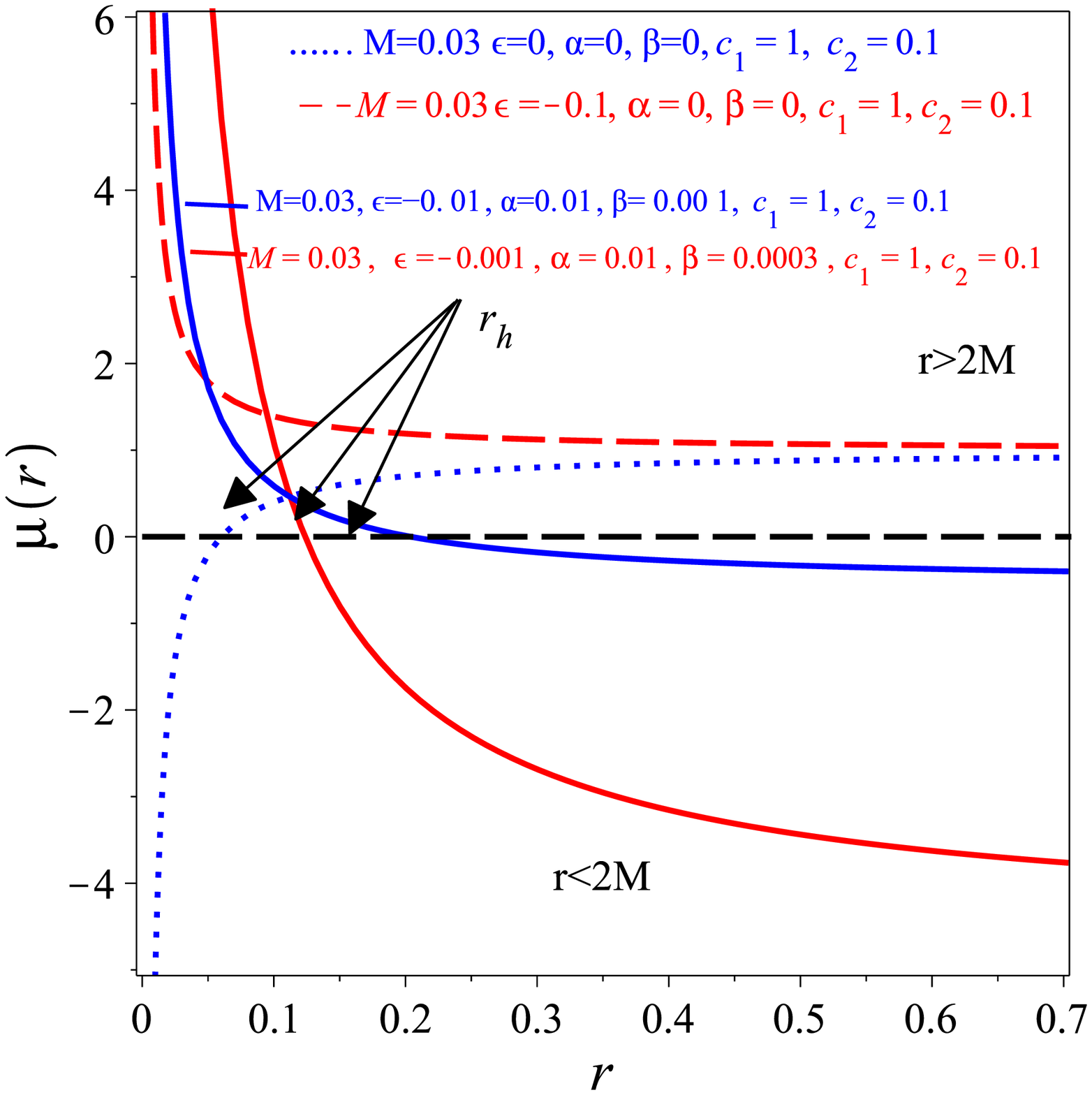}}
\subfigure[~The horizon mass-radius]{\label{fig:6b}\includegraphics[scale=0.3]{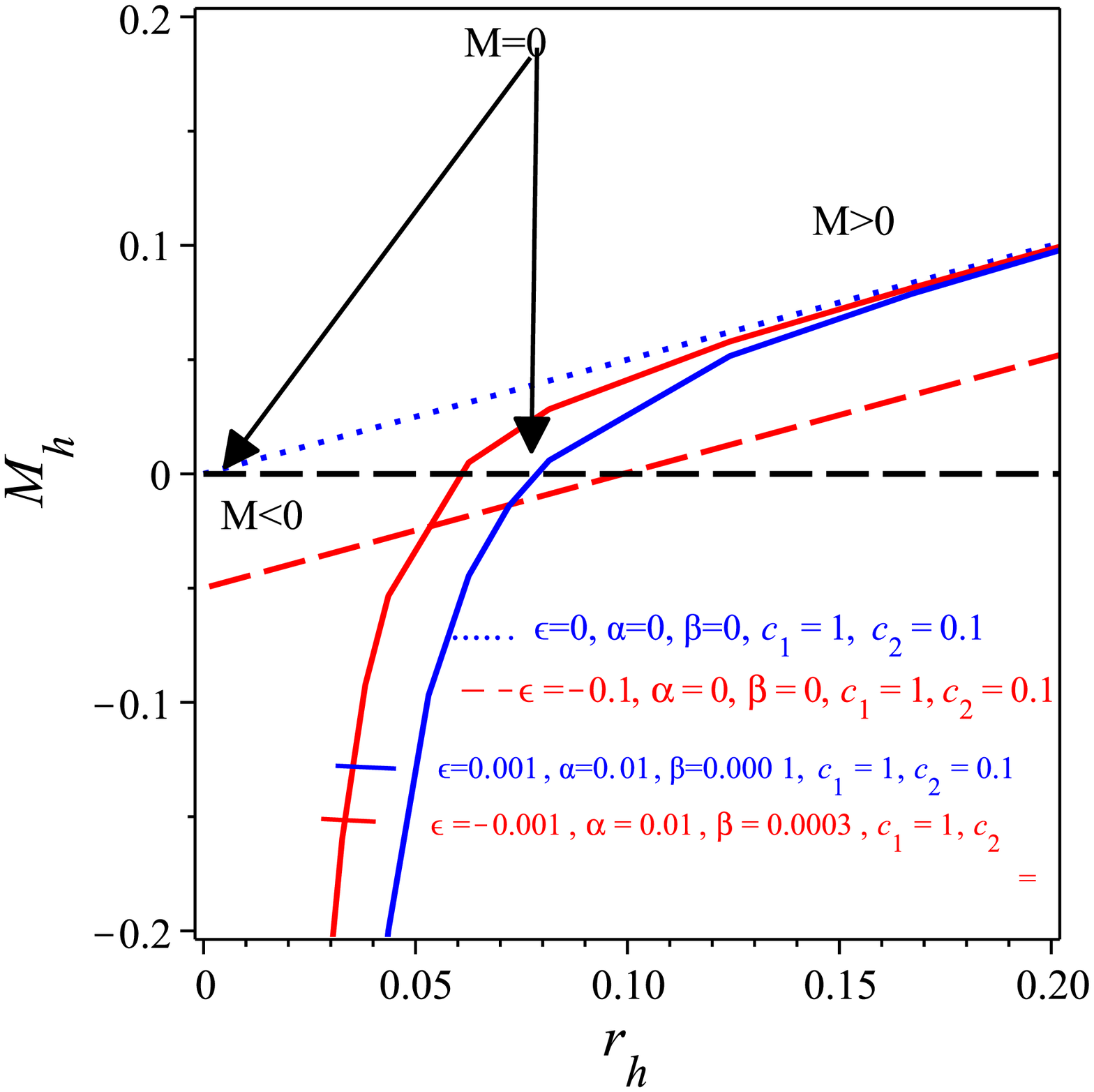}}
\subfigure[~The horizon Bekenstein-Hawking entropy]{\label{fig:6c}\includegraphics[scale=0.3]{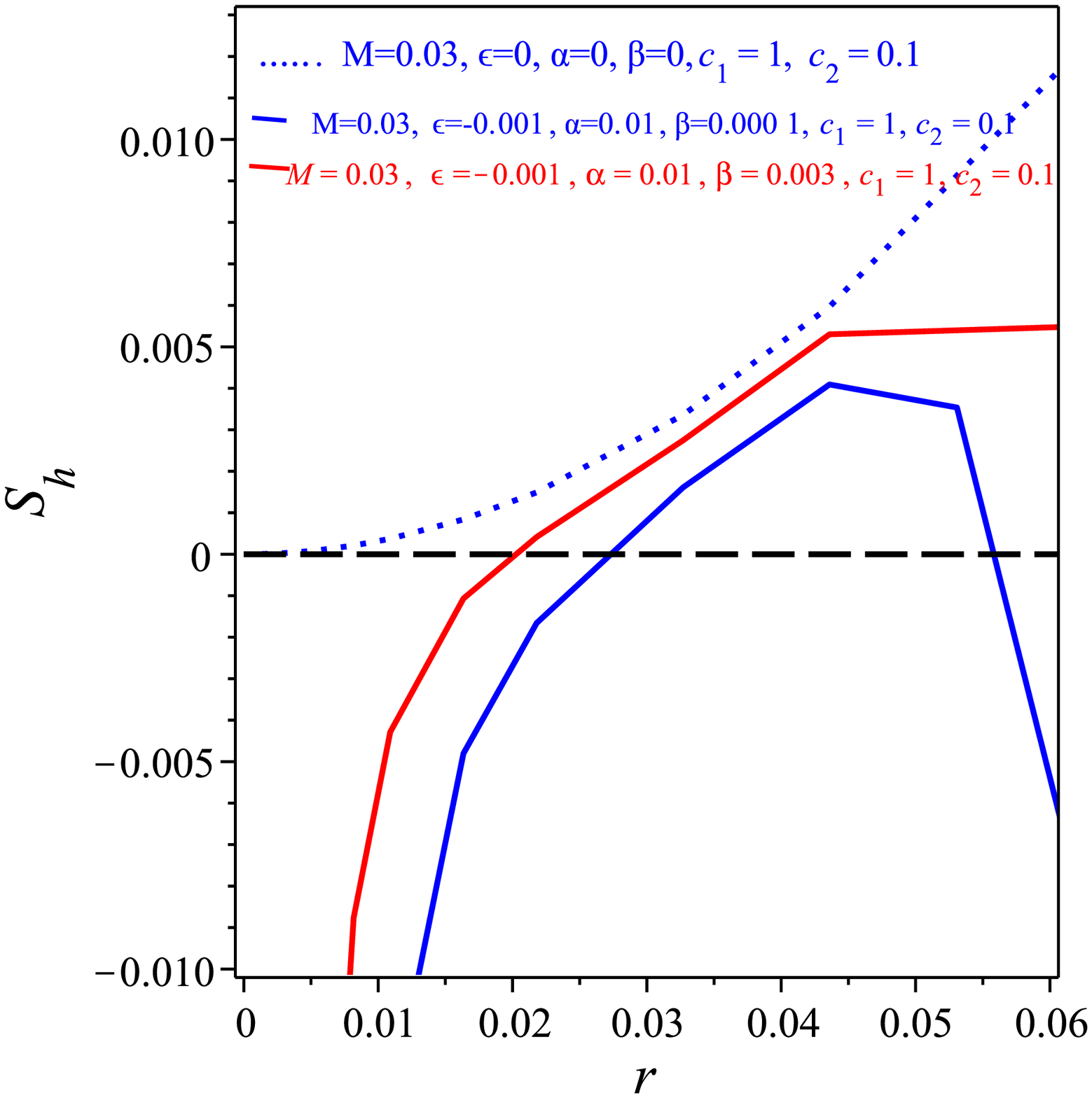}}
\subfigure[~The horizon heat capacity]{\label{fig:6d}\includegraphics[scale=0.3]{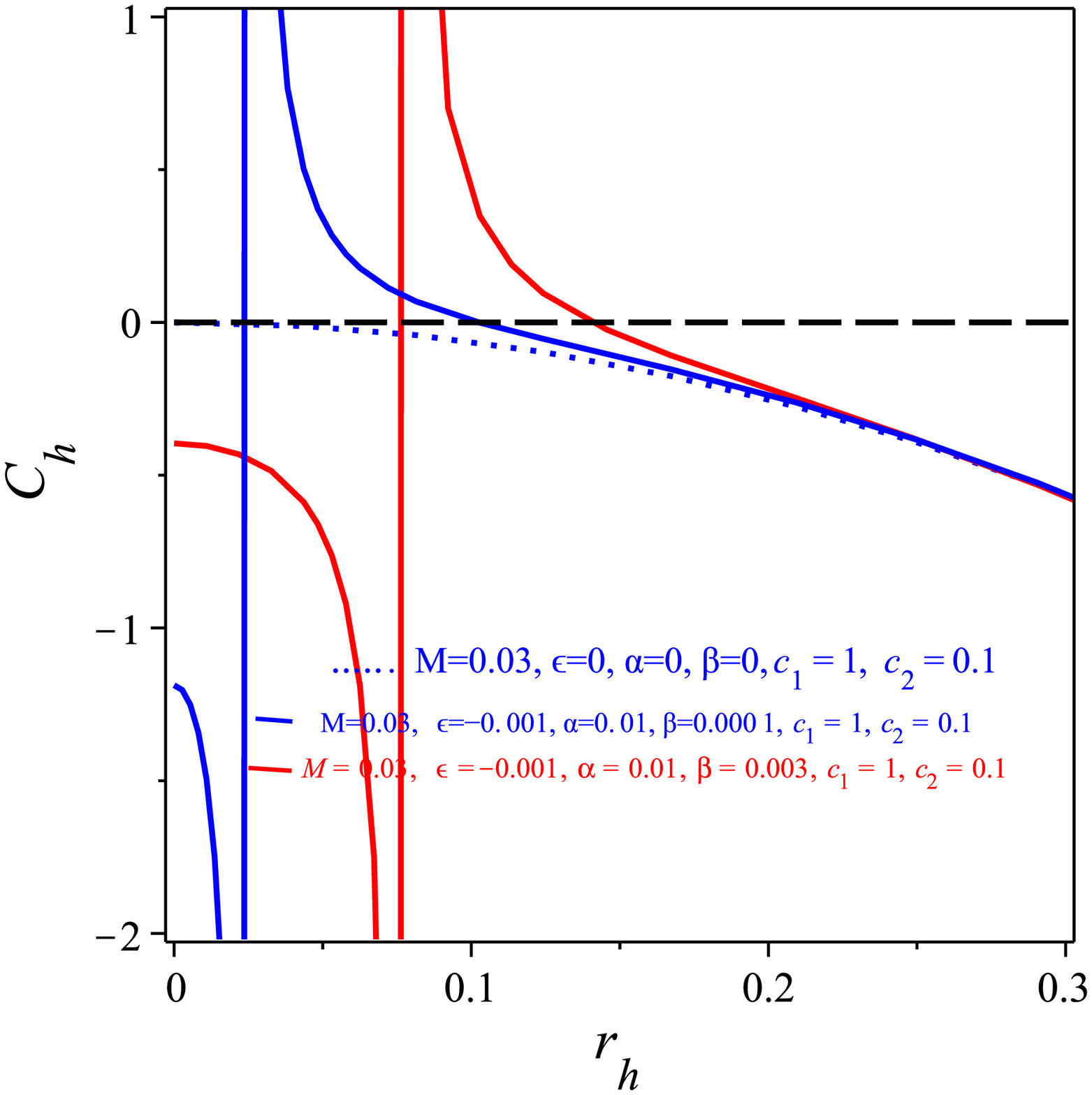}}
\subfigure[~The horizon Gibbs free energy]{\label{fig:6e}\includegraphics[scale=0.3]{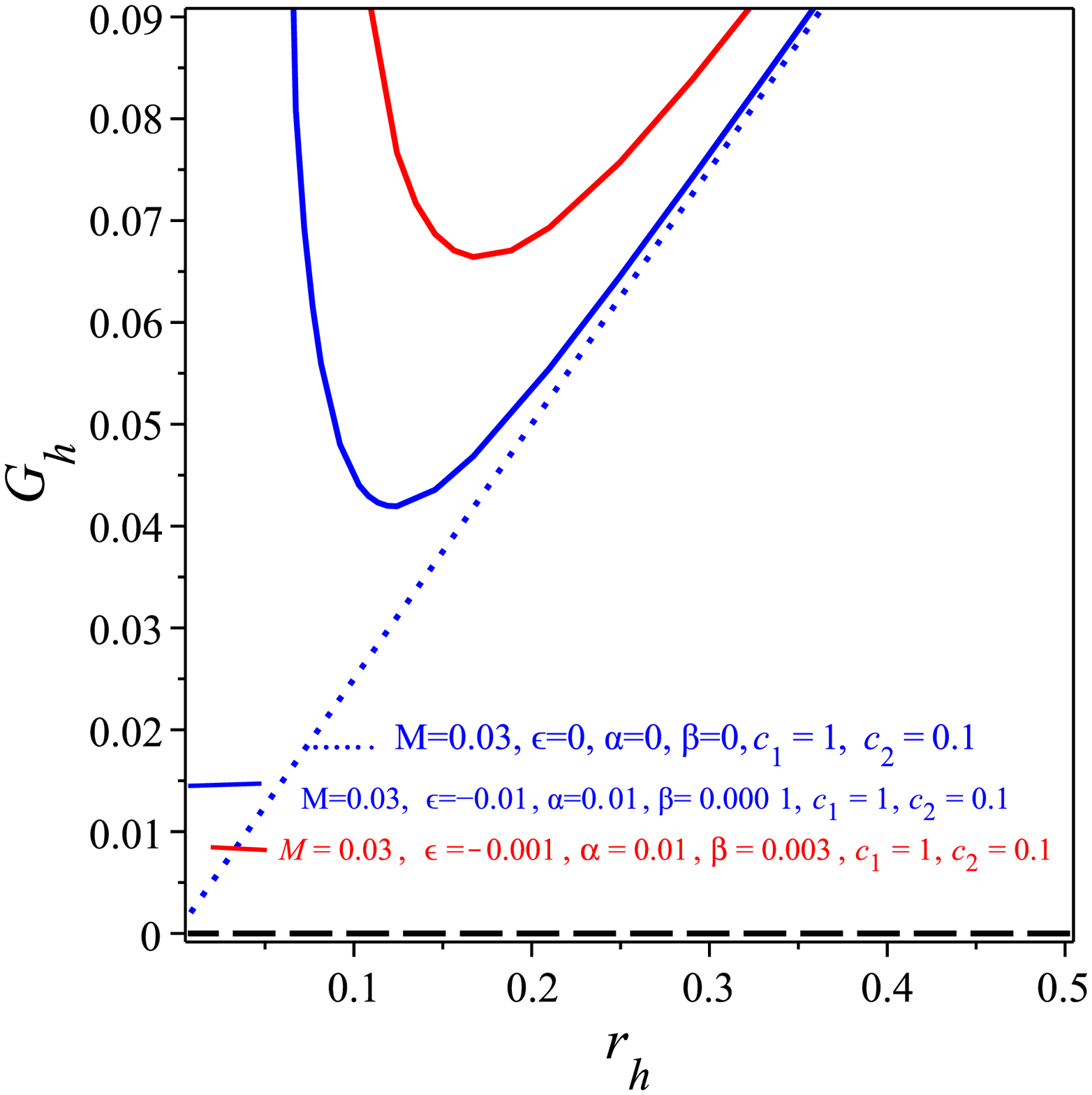}}
\caption{Schematic plots of thermodynamical quantities of the black hole solution (\ref{pot}) for negative value of $\epsilon$: \subref{fig:6a}
 Typical behavior of the metric function $\mu(r)$ given by (\ref{pot}); \subref{fig:6b} the horizon mass--radius relation (\ref{hor-mass-rad1a});
  \subref{fig:6c}
   typical behavior of the horizon entropy, which shows that $S_h$ increases quadratically as $r_h$ increases and then decrease; \subref{fig:6d}
    typical behavior of  the heat capacity, (\ref{heat-cap1ac}),
     which shows that it vanish at $r_{h}$, and
    as $r<r_h$ we have a
   positive  heat capacity and as $r>r_h$ we have a negative value of
the black hole;
     \subref{fig:6e} typical behavior of the horizon Gibbs free energy which shows that $G_h$ have always
      positive value for negative value of $\epsilon$.}
\label{Fig:6}
\end{figure}

  From Eqs. (\ref{hor-mass-rad1ac}),
(\ref{ent1c}) and (\ref{m44ec}) in (\ref{enr1}), we obtain
\begin{eqnarray} \label{m77c}
&&{G_h}_{{}_{{}_{{}_{{}_{cubic}}}}}=\frac{\pi}{540r_h{}^3(r_h{}^2-4m^2)^2}\Bigg[135r_h{}^4(r_h{}^2-4m^2)^2-\epsilon\Bigg\{4(1575r_h{}^6\alpha+
9272r_h{}^4\beta+133232\beta m^4)-135r_h{}^7[r_hc_2-2c_1]\nonumber\\
&&-16r_h m^2\beta[17329 r_h-270m]+1080c_2m^2r_h{}^4[r_h{}^2-2m^2]-2160m^2r_h{}^3c_1[r_h{}^2-2m^2]
+405r_h{}^6\alpha ln(2)\nonumber\\
&&+\Big\{2160r_h m^2[4\beta +\alpha r_h{}^2]+540r_h{}^2\alpha m[r_h{}^2 -4m^2]-4320
\beta m[2m^2 -4r_h{}^2]-540r_h{}^3[8\beta+r_h{}^2\alpha]\Big\}\sqrt{r_h{}^2+2mr_h}\nonumber\\
&&-10r_h{}^2m^2\alpha[4932r_h{}^2-9648m^2-648m^2ln(2)
+324r_h{}^2ln(2)]\Bigg\}\Bigg].
\end{eqnarray}
We depict  Gibbs energy of the black hole (\ref{pot}) in Fig.  \ref{Fig:5}\subref{fig:5f},   which indicates that the Gibbs energy has negative values
 at large $r<r_h$ and positive value when $r>r_h$. We note that for $\epsilon=0$, the Scharzschild black hole is recovered which is shown in Fig.
  \ref{Fig:5}\subref{fig:5f} by the blue dot curve. Interestingly, for negative value
of $\epsilon$, Gibbs energy is always positive and as we discuss before that the price of this is the negative Hawking temperature. We depict the case of negative
value of $\epsilon$ in Fig. \ref{Fig:6}.
\newpage
We summarize the results of thermodynamics of the quadratic and cubic cases in table I. As this table shows that when the parameter $\epsilon$ takes positive value we have   good models.

\begin{table*}[t!]
   \caption{\label{Table1}%
The main results thermodynamic  according to the values of $\epsilon$, i.e., positive or negative values.}
\begin{ruledtabular}
\begin{tabular*}{\textwidth}{lcccccc}
&&&\bf{Quadratic}  &&\\
\hline
{$\epsilon$}                       & Entropy& Temperature  && Heat capacity &Gibb's free energy          \\
\hline
zero                        &positive& positive&&negative& positive\\
$>$zero                        &positive& positive&&negative& positive(conditional)\\
$ <$zero                        &positive(conditional)& positive(conditional)&&positive(conditional)& positive\\
\hline
 & &&\bf {Cubic}     & &          \\
\hline
zero                        &positive& positive&&negative& positive\\
$>$zero                        &positive& positive&&negative& positive(conditional)\\
$<$zero                        &positive(conditional)& positive(conditional)&&positive(conditional)& positive\\
\end{tabular*}
\end{ruledtabular}
\end{table*}
 \section{Discussion and conclusions}\label{S6}

In this study, we further studied  $f(T)$ in the framework of the spherically symmetric space-time. To do this, we used a physical tetrad field space-time  possessing  two unknown functions. This study  aims to determine  the physics of the BH that was derived in \cite{Bahamonde:2019zea}
for the quadratic form of $f(T)$. We calculated the invariants of this BH and demonstrated that the behavior of the tensors $T^{\alpha \beta \gamma}
T_{\alpha \beta \gamma}=T^{\alpha}T_{\alpha}=T\sim  {r^{-2}}$  are similar to the results obtained before in the frame of $f(T)$ \cite{Nashed:2018cth,2017JHEP...07..136A}, and  dissimilar  to the  solutions of  GR and TEGR, which behaved as $T^{\alpha \beta \gamma}
T_{\alpha \beta \gamma}=T^{\alpha}T_{\alpha}=T\sim  {r^{-4}}$. This  clearly indicated that the contribution of the higher-order torsion made the
 singularity of  $T^{\alpha \beta \gamma}
T_{\alpha \beta \gamma}$, $T^{\alpha}T_{\alpha}$  and $T$ milder. Additionally, we calculated the energy content and revealed  the
  contribution of the higher-order torsion. Finally, we calculated the stability of this BH via the geodesic deviation and revealed the regions
   where the BH exhibited  stability, as shown in Fig. \ref{Fig:1}.

 In the second part, we presented the cubic form of the field equations using $f(T)=T+\epsilon\Big[\frac{1}{2}\alpha T^2+\frac{1}{3}\beta T^3\Big]$
  and derived the asymptotic form of these field equations up to $O(\epsilon)$. The derived BH solution was characterized by the mass, the
  two constants of integrations and the three constants that characterized the form  $f(T)$. The asymptotic form of this BH behaved like a flat
  space-time. We calculated the invariants of this BH and its energy thereby revealed the contribution of the higher-order torsion to the energy. Furthermore,
  we calculated the geodesic deviation and revealed where the regions of stability in the BH.  To study the BH solutions of this study in-depth we calculated the thermodynamic quantities including the Hawking temperature, entropy,  heat capacity, and Gibbs
 energy for the quadratic and cubic forms of the BH solutions. For the quadratic form, we used the same values of the constants that are used for the stability study. We studied all the thermodynamical quantities with two values of the parameters $\epsilon$ the positive value and negative values as shown in
 figures \ref{Fig:3} and \ref{Fig:4}. The most beneficial  results that were discussed were those of the  entropy, Hawking temperature, heat capacity, and Gibbs energy. At $\epsilon>0$  positive values were always obtained for the entropy and temperature while  a negative value was obtained for the heat capacity. Regarding the  Gibbs energy, we
  obtained a negative value at  $r_{dg}<r_h$ and a positive one  at $r_{dg}>r_h$. At $\epsilon <0$ those quantities were negative regarding the entropy
   at $r_{dg}<r_h$ and positive at $r_{dg}>r_h$. Furthermore, regarding  the temperature and heat capacity the values were negative at $r_{dg}<r_h$ and positive
  at $r_{dg}>r_h$. Regarding the Gibbs energy it was always  positive and we concluded that $\epsilon$ of the quadratic form
   might be positive or negative values and both of them afforded the physical thermodynamic quantities.   The same discussion is applicable for the cubic BH.

%

\end{document}